\documentclass[twocolumn,showpacs,preprintnumbers,amsmath,amssymb,latexsym,prd,nofootinbib,floatfix]{revtex4-1}
\usepackage{natbib}
\usepackage{amsmath}
\usepackage{graphicx}
\usepackage{relsize}
\usepackage{MnSymbol}
\usepackage{xcolor}
\usepackage[utf8]{inputenc}

\begin{document}

\title{Reliable estimation of the radius of convergence in finite density QCD}
\author{M. Giordano}
\author{A. P\'asztor}
\affiliation{ELTE E\"otv\"os Lor\'and University, Institute for
  Theoretical Physics, P\'azm\'any P.\ s.\ 1/A, H-1117, Budapest,
  Hungary}

\begin{abstract}
We study different estimators of the radius of convergence of 
the Taylor series of the pressure in finite density QCD. 
We adopt the approach in which the radius of convergence is estimated
first in a finite volume, and the infinite-volume limit is taken later. 
This requires an estimator for the radius of convergence that is
reliable in a finite volume. Based on general arguments about the
analytic structure of the partition function in a finite volume, we
demonstrate that the ratio estimator cannot work in this approach, and
propose three new 
estimators, capable of extracting reliably the radius of convergence,
which coincides with the distance from the origin of the closest
Lee-Yang zero. We also provide an estimator for the phase of the
closest Lee-Yang zero, necessary to assess whether the leading
singularity is a true critical point. We demonstrate the usage of
these estimators on a toy model, namely 4 flavors of unimproved
staggered fermions on a small $6^3 \times 4$ lattice, where both the
radius of convergence and the Taylor coefficients to any order can be
obtained by a direct determination of the Lee-Yang zeros.
Interestingly, while the relative statistical error of the Taylor
expansion coefficients steadily grows with order, that of our
estimators stabilizes, allowing for an accurate estimate of the radius
of convergence. 
In particular, we show that despite the more than 100\% error bars on
high-order Taylor coefficients, the given ensemble contains enough
information about the radius of convergence. 
\end{abstract}

\maketitle

\section{Introduction}
One of the most important unsolved problems in QCD is the
determination of its phase diagram at finite baryonic density. In
particular, an open question is whether the analytic crossover of the
chiral transition at zero chemical potential turns into a genuine
phase transition sufficiently deep in the $\mu-T$ (chemical
potential-temperature) plane, and, if so, where the critical
endpoint lies. Nonperturbative studies of these questions on the
lattice are notoriously hampered by the sign problem, which prevents
the use of standard Monte Carlo techniques to probe QCD directly at
finite density. Reweighting
techniques~\cite{Hasenfratz:1991ax,Barbour:1997ej,Fodor:2001au, 
Fodor:2001pe,Fodor:2004nz,Csikor:2004ik} shift 
this problem from the integration measure to the observable, allowing
the use of standard Monte Carlo techniques, but they are still limited
in scope by the sign and overlap problems, both of which are
exponentially hard in the lattice volume.
At the moment, the state of the art for lattices close to the
continuum limit is to calculate Taylor coefficients of the pressure,
either by direct calculations at $\mu=0$
~\cite{Allton:2002zi,Allton:2005gk,Gavai:2008zr,
Basak:2009uv,Borsanyi:2011sw,Borsanyi:2012cr,Bellwied:2015rza,
Ding:2015fca,Bazavov:2017dus}  
or via fitting a polynomial ansatz to results obtained with
simulations at zero and imaginary chemical
potentials~\cite{deForcrand:2002hgr,DElia:2002tig,DElia:2009pdy, 
Bonati:2014kpa,Bonati:2015bha,DElia:2016jqh,Gunther:2016vcp,
Vovchenko:2017xad,Bonati:2018nut,Borsanyi:2018grb}.   
When trying to infer if the calculated Taylor coefficients show any sign of
criticality, a common choice is to use simple estimators of the radius
of convergence, like the ratio
estimator~\cite{DElia:2016jqh,Bazavov:2017dus,Fodor:2018wul}.  
The purpose of this paper is to obtain some insight on the usefulness
of such estimators, using both 
analytical and numerical methods. 

The study of the radius of convergence $r$ from the Taylor coefficients 
can be carried out in two different ways. One is to perform first the
infinite volume extrapolation of every coefficient, and in the next
step estimate the radius of convergence from the large-order limit of
an appropriate estimator. In other words, given an estimator $R_n(V)$
of $r$, where $V$ is the volume of the system and $n$ is the order of
the Taylor expansion, one determines $r$ by taking limits in the order
$r =\lim_{n \to \infty} \lim_{V \to \infty} R_n(V)$. The other
procedure is to estimate the radius of convergence in a finite volume
first, and in the next step do the infinite volume extrapolation,
i.e., taking limits in the order $r = \lim_{V \to \infty} \lim_{n \to
  \infty} R_n(V)$. While both approaches are valid and worth being
investigated, in this paper we will focus on the latter, mainly for
the following reasons. 

First, in a finite volume, the analytic form of the large-order
behavior of the Taylor expansion of the pressure is particularly
simple, allowing us to produce 
analytical arguments about the accuracy
and reliability of different convergence radius estimators, which we
will proceed to do in this paper. 

Second, the infinite-volume extrapolation in finite density QCD is
very hard, mainly because the sign problem is exponentially hard in
the volume. Any fixed-order calculation of the Taylor coefficients has
only polynomial complexity in the volume, but the order of the
polynomial increases with the order of the coefficient, eventually
recovering the full exponential complexity in the infinite-order
limit. This makes it useful to study methods that can give
well defined results already in a finite volume. 

In a finite volume, the radius of convergence coincides with the 
distance from the origin of the closest Lee-Yang
zero~\cite{Lee:1952ig}, i.e.,  
the zero of the partition function closest to the origin in the
complex $\mu$ plane. 
The behavior of Lee-Yang zeros in the thermodynamic limit
provides very useful information about the phase diagram of the
theory. In fact, the presence of a critical point is signaled by 
Lee-Yang zeros reaching somewhere the real $\mu$ axis in the
infinite-volume limit. The rate at which the real axis is approached
tells us about the nature and the strength of the transition: in
particular, a first order phase transition is signaled by the
imaginary part of the nearby Lee-Yang zeros vanishing like $1/V$. The
large-volume limit of the real part of such zeros determines of course
the location of the critical point. Zeros that even in the
thermodynamic limit remain a finite but small distance away from the
real axis are expected to correspond to an analytic crossover.
It is clear that while the Lee-Yang zeros determine
both the critical points (if any) of the theory and the radius of
convergence of the Taylor series of the pressure, nothing guarantees
that the same zeros are involved in the two cases. Nonetheless, this
is entirely possible, and in the worst case the radius of convergence
provides a lower bound on the critical endpoint. 

In this paper 
we study whether one can accurately infer the position of the 
leading Lee-Yang zero from the Taylor coefficients of the pressure,
assessing the reliability of the various radius estimators and their 
computational cost. In Section \ref{sec:taylor} we discuss
Lee-Yang zeros in QCD in some detail, relate them to the Taylor
coefficients of the pressure, and exploit their symmetries to assess
the viability of the ratio estimator, and of three new proposals to
estimate the radius of convergence. We also discuss Fisher zeros,
i.e., zeros of the partition function in the complex gauge coupling
plane, and show how they relate to the cumulants of the gauge
action. 
In Section \ref{sec:unimp} we test
our arguments by computing all the Lee-Yang zeros in a toy model for
lattice QCD, 
namely $N_f=4$ unimproved 
unrooted staggered fermions on  a small, $6^3\times 4$ lattice, and
comparing the radius of convergence 
extracted from the Taylor coefficients to the known position of the
closest Lee-Yang zero. 
To further test our methods, we apply our convergence radius
estimators to find the Fisher zero closest to the real axis, and
compare it with a direct determination using reweighting. 
Our conclusions are presented in Section
\ref{sec:concl}.

\section{Lee-Yang zeros, Taylor series in a finite volume, and
  convergence radius estimators}
\label{sec:taylor}
We will study the convergence of several different Taylor series 
expansions of the pressure $p = \frac{T}{V} \log Z$, where $Z$ is the grand 
canonical partition function.
For a relativistic theory on the lattice in a finite volume, $Z$ reads:
\begin{equation}
\label{eq:part_func}
    Z(\hat{\mu}) = \mathlarger{\sum}_{n=-k V}^{+k V} Z_n e^{n
      \hat{\mu}} \,. 
\end{equation}
Here $\hat{\mu}=\mu/T$,  $V=N_s^3$ is the lattice volume, $Z_n$ are 
temperature-dependent real positive coefficients, 
and $k$ is a constant depending on the particular model. 
The coefficients $Z_n$ are the canonical partition
functions, corresponding to the sector of Hilbert space where the
conserved charge conjugated to $\mu$ equals $n$. In the case of the
model to be discussed below in Section \ref{sec:unimp}, i.e., $N_f=4$
lattice QCD with unrooted staggered fermions, we have $k=3$. 

The partition function is a non-vanishing analytic 
function of $\hat{\mu}$ (i.e., $e^{-kV\hat{\mu}}$) times a polynomial
of order $2kV$ in the fugacity $e^{\hat{\mu}}$, and by the fundamental
theorem of algebra it has $2k V$ roots in the complex fugacity plane,
called Lee-Yang zeros~\cite{Lee:1952ig}.  
In terms of the Lee-Yang zeros the 
partition function can be written as follows, 
\begin{equation}
    Z(\hat{\mu}) = 
\text{non-vanishing\ } \cdot  \prod_{m=1}^{2k V}
    (\Lambda_m-e^{\hat{\mu}}) \,.
\end{equation}
The logarithm of the partition function therefore has the form:
\begin{equation}
\label{eq:sum_fug}
\frac{pV}{T}=     \log Z(\hat{\mu}) =  
\text{analytic\ } + \sum_{m=1}^{2k V}   
\log   (\Lambda_m-e^{\hat{\mu}})\,.
\end{equation}
Since the finite-volume partition function as a function of $\hat{\mu}$ is a
finite sum of exponentials, and so an entire function, the 
Weierstrass factorization theorem (see, e.g.,
Ref.~\cite{Rudin_analysis}) states that a similar factorization also
exists in terms of $\hat{\mu}$, except that 
now one has an infinite product. As a function of $\hat{\mu}$,
$\log Z$ has therefore the same form as
Eq.~\eqref{eq:sum_fug}, with the finite sum replaced by an infinite one:
\begin{equation}
\label{eq:sum_mu}
    \log Z(\hat{\mu}) =  
\text{analytic\ } + \sum_{m=1}^{\infty} \log
    (\lambda_m-\hat{\mu})\,. 
\end{equation}
This can be easily understood by noticing that to each $\Lambda_m$
corresponds an infinite tower of Lee-Yang zeros $\lambda_{m,n} = {\rm Log}\,
\Lambda_m  + 2\pi n i $, where ${\rm Log}$ is the principal branch of the
complex logarithm. Explicitly, to each term in the expansion in fugacity
correspond the following terms in the expansion in $\hat{\mu}$:
\begin{equation}
\begin{aligned}
\log \left( e^{\hat{\mu}} - \Lambda \right) &= \frac{\hat{\mu}}{2} +
\frac{1}{2} {\rm Log} \Lambda + {\rm Log} \left( \hat{\mu} - \rm{Log}
  \Lambda \right) + \\ 
&+ \sum_{k=1}^{\infty} \left[ \log(\hat{\mu}-\rm{Log} \Lambda + i 2 \pi k ) -
  \log(2 \pi k) \right] \\  
&+ \sum_{k=1}^{\infty} \left[ \log(\hat{\mu}-\rm{Log} \Lambda - i 2 \pi k) -
  \log(2 \pi k) \right] \,,
\end{aligned}
\end{equation}
as can be easily worked out by noticing that
$e^{\hat{\mu}} - \Lambda = 2 e^{\hat{\mu}/2} \sqrt{\Lambda}
\sinh(\frac{\hat{\mu}}{2}-\frac{1}{2} \rm{Log} \Lambda )$ and
utilizing $\sinh(x) = x \prod_{k=1}^{\infty}\left( 1 +
  \frac{x^2}{\pi^2 k^2}\right)$. 

The large-order behavior of the Taylor series of $\log Z$  
will be determined by the location of the logarithmic singularity
closest to the expansion point in Eqs.~\eqref{eq:sum_fug} or
\eqref{eq:sum_mu}. The relevant choices are here $e^{\hat{\mu}}=1$ or
$\hat{\mu}=0$, respectively. The two types of expansion
can be treated similarly. 
In both cases the relevant contributions to
the partition function have the Taylor series  
\begin{equation}
\label{eq:basic_term}
    \log(A-x) = \log A + \sum_{k=1}^{\infty}\frac{-1}{k} \frac{1}{A^k} x^k = 
		b_0 + \sum_{k=1}^{\infty} b_k x^k\,,
\end{equation}
where the expansion parameter $x$ is either the ``fugacity parameter''
$\zeta \equiv e^{\hat{\mu}}-1$ or the chemical potential over
temperature $\hat{\mu}$, and the relevant Lee-Yang zero $A$ is 
correspondingly $\Lambda_1-1$ or $\lambda_1$, having assigned the
index $m=1$ to the leading singularity.

The discussion above is rather general. In the specific case of QCD,
on which we will be focussing our attention from now on,
the symmetries of the theory imply useful relations for the
grand canonical partition function $Z(\hat{\mu})$ and for the canonical
partition functions $Z_n$. Due to CP symmetry (at zero
$\theta$-angle), one finds $Z(-\hat{\mu})=Z(\hat{\mu})$ (for complex $\hat{\mu}$),
and $Z_{-n}=Z_n$. As we have already mentioned above, the $Z_n$ are
real, which implies that $Z$ is a real analytic function,
$Z(\hat{\mu}^*)=Z(\hat{\mu})^*$. It is easy to see that $Z$ is $2\pi$-periodic in
the imaginary direction, $Z(\hat{\mu}+i2\pi)=Z(\hat{\mu})$, and furthermore,
thanks to the Roberge-Weiss symmetry~\cite{Roberge:1986mm}, one finds
$Z(\hat{\mu}+i2\pi/3)=Z(\hat{\mu})$. 
This implies that the $Z_n$ vanish if $n$ is
not a multiple of 3. Combining this with the $\hat{\mu} \to -\hat{\mu}$ symmetry,
one also concludes that $Z(\hat{\mu}+i\pi/3)=Z(-\hat{\mu}+i\pi/3)$. These
properties imply useful symmetries for the distribution of Lee-Yang
zeros in the complex plane. In the complex fugacity plane, such a
distribution is invariant under reflection through 
the real
axis ($\Lambda\to\Lambda^*$), inversion with respect to the unit
circle ($\Lambda\to1/\Lambda$), and rotations of multiples of
$\frac{2\pi}{3}$ ($\Lambda\to e^{i\frac{2\pi}{3}}\Lambda$).
It is thus sufficient to focus on the region $\{ z = r e^{i\theta} |~ r
\le 1\,, ~~ 0\le \theta \le \pi/3\}$ in the complex fugacity plane to
determine completely the Lee-Yang zeros of the partition function. 
In general, due to the positivity of the canonical partition functions
$Z_n$, no zeros will be found on the real $\hat{\mu}$ axis, and so on the
real positive fugacity axis. In the case of QCD with staggered
fermions, the fermionic determinant appearing in the
functional-integral representation of the partition function is
positive definite for purely imaginary $\hat{\mu}$, and so no zeros will
appear on the imaginary $\hat{\mu}$ axis, or equivalently on the unit circle
in the complex fugacity plane.

\subsection{The failure of the ratio estimator}

A common estimator for the radius of convergence $r$ of a power series 
$f(x) = \sum_k B_k x^k$ is the ratio estimator $R_k \equiv |B_k/B_{k+1}|$,
from which one obtains $r$ as $r=\lim_{k\to\infty} R_k$ when this
limit exists. 
In the case of the QCD partition function, one naively expects that in
the limit of large order 
$R_k$
will be dominated by a single term like that in
Eq.~\eqref{eq:basic_term}, thus converging to 
\begin{equation}
  \label{eq:naive_est}
R_k \to \left| \frac{b_k}{b_{k+1}} \right| = \left|\frac{(k+1)A^{k+1}}{k A^k}\right| \mathop\to_{k\to\infty} |A| \,,
\end{equation}
as $k\to\infty$. However, a more careful analysis of the
analytic structure of the partition function shows that this is not
the case. Let us look first at the case of the expansion in the
fugacity parameter $\zeta=e^{\hat{\mu}}-1$. 
Since 
$Z(\hat{\mu}^*)=Z(\hat{\mu})^*$ and $\zeta(\hat{\mu}^*)=\zeta(\hat{\mu})^*$, 
the Lee-Yang zeros come in complex-conjugate pairs. There are then two
leading singularities, giving the following contributions to the
partition function, 
\begin{equation}
    \label{eq:taylor_1ly}
\begin{aligned}
&    \log(A-\zeta) + \log(A^*-\zeta)      \\
&=		\log A +     \log A^* + 
\sum_{k=1}^{\infty}\frac{-2}{k}
    \frac{\cos(k \theta)}{r^k} \zeta^k  \\
&=		c^{(\zeta)}_0 +\sum_{k=1}^{\infty} c^{(\zeta)}_k \zeta^k
\,,
\end{aligned}
\end{equation}
where $r$ and $\theta$ are the polar coordinates of $A=\Lambda_1-1 =
re^{i\theta}$ in the complex plane. Here it is understood
that $|\Lambda_1|<1$.\footnote{It is straightforward to show that for the
  corresponding Lee-Yang zeros outside the unit circle  $$|1/\Lambda_1
  -1 |=|1/\Lambda_1^* -1 | > |A|\,.$$} 
The ratio estimators are then:
\begin{equation}
    R_k^{(\zeta)} = \left| \frac{c^{(\zeta)}_k}{c^{(\zeta)}_{k+1}}
    \right| = r \,\frac{k+1}{k}  \left| 
      \frac{\cos(k \theta)}{\cos((k+1) \theta)} \right| \,.
\end{equation}
Due to the last factor in this expression, the ratio estimators will
never converge in a finite volume for $\theta\neq 0,\pm \pi/2,\pi$. Of course,
in a finite volume we never have $\theta = 0$ or $\pi$, due to the strict
positivity of the partition function.
Even in an infinite volume the ratio estimators will not
converge, unless $\theta$ converges to one of the values $0,\pm
\pi/2,\pi$. Luckily enough, these include the physically interesting
case of a genuine phase transition (in this case corresponding to
$\theta=\pi$ and $r<1$).
On the other hand, the ratio estimator will not work for a
general crossover transition corresponding to a complex singularity even 
in the thermodynamic limit (except for the very special case $\theta=\pm
\pi/2$). In this case, even if a genuine phase transition at real
$\mu$ exists, but is not the closest singularity to $\zeta=0$, the
ratio estimators will not converge, and so will not even give a lower
limit on the location of the critical point. 

The situation is very similar when expanding in $\hat{\mu}$, except 
that, since $Z(\hat{\mu})=Z(-\hat{\mu})$, we now have to take into account four
Lee-Yang zeros at the same distance from the 
expansion point, so that the Taylor expansion reads
\begin{equation}
  \begin{aligned}
    \label{eq:taylor_1ly_mu}
&    \log(A-\hat{\mu}) + \log(A^*-\hat{\mu}) + \log(-A-\hat{\mu}) +
\log(-A^*-\hat{\mu})  
\\ &=    \log A + \log A^* + \log(-A) + \log(-A^*)
+ \\
&+   \sum_{k=1}^{\infty}\frac{-2}{k} \frac{\cos(k \theta)+\cos(k(\pi -
      \theta))}{r^k} \hat{\mu}^k\\
&= c^{(\hat{\mu})}_0 +    \sum_{k=1}^{\infty}c^{(\hat{\mu})}_k \hat{\mu}^{2k}\,.
  \end{aligned}
\end{equation}
Here $A=\lambda_1=re^{i\theta}$, and
it is understood that $0\le \theta\le\frac{\pi}{2}$.
The expansion coefficients of course vanish for odd order,
corresponding to the fact that the pressure is an even function of the 
chemical potential. The ratio estimators read in this case
\begin{equation}
    \left(R_{k}^{(\hat{\mu})}\right)^2 = 
\left| \frac{c^{(\hat{\mu})}_{k}}{c^{(\hat{\mu})}_{k+1}} \right| =
  r^2\,  \frac{k+1}{k}  \left| \frac{\cos(2k \theta)}{\cos(2(k+1) \theta)}
    \right| \,.
\end{equation}
This is essentially the same formula as for the fugacity parameter, up
to the substitutions $r\to r^2$ and  $\theta \to 2\theta$, and
therefore exactly the same comments apply. In particular, the ratio
estimator will work only if $\theta=0,\frac{\pi}{4},\frac{\pi}{2}$, which
include (in the thermodynamic limit) genuine phase transitions at real or purely imaginary $\mu$,
of course in case that they correspond to the closest singularity. 

The similarity between the two cases is made even clearer if one
realizes that $Z(\hat{\mu})$ is, in fact, an entire function of
$\hat{\mu}^2$, as a consequence of the symmetry $Z_n = Z_{-n}$ of the
QCD canonical partition functions. One can then repeat the discussion
in terms of $\hat{\mu}^2$, starting from Weierstrass factorization and
noticing that the roots are now nothing but $\lambda_m^2$, and come in
complex conjugate pairs. This leads directly to
Eq.~\eqref{eq:taylor_1ly_mu} for the contribution of the leading pair 
of singularities, which is manifestly an expansion in
$\hat{\mu}^2$. It is evident that the Taylor coefficients 
$c_k^{(\zeta)}$ and $c_k^{(\hat{\mu})}$ have the same form when the
latter are expressed in terms of the polar coordinates of the
leading zero in $\hat{\mu}^2$, i.e., $\lambda_1^2$, which is
understood to lie in the upper half of the complex
$\hat{\mu}^2$-plane.

One final comment: the leading Lee-Yang zero for the two expansions
need not be the same, i.e., in general $e^{\lambda_1} \neq \Lambda_1$, 
since changing the expansion parameter corresponds to a conformal map 
on the complex chemical potential plane, that can change the ordering
of the distances of the singularities from the origin. Such a situation 
of changing the ordering of the singularities was pointed
out for the example of a chiral effective model in Ref.~\cite{Skokov:2010uc}.

\subsection{Fisher zeros and cumulants of the gauge action}

A rather similar argument applies if, instead of the
chemical potential $\hat{\mu}$, one considers the gauge coupling
$\beta$ as a complex variable. One starts by writing the lattice
partition function in its path-integral representation,
\begin{equation}
\label{eq:path_integral}
Z(\beta)= \int \mathcal{D}U e^{-\beta G} \det M\,,
\end{equation}
where $\mathcal{D}U$ denotes invariant group integration of the gauge
links, $G$ is the $\beta$-independent gauge action, and $M$ is the
fermion matrix. 
For a finite number of SU(3) integrals $Z$ is an entire function of
the gauge coupling $\beta$. 
Complexifying the coupling as $\beta \to \beta_C= \beta + w$ with
$\beta$ kept real, having in mind to perform simulations at $\beta$,
one can exploit Weierstrass factorization to write
\begin{equation}
\label{eq:weier_beta}
Z(\beta+w) \equiv f(w)= f(0) e^{w\frac{f'(0)}{f(0)}} \prod_k \left[
  \left( 1- \frac{w}{w_k}\right) e^{w/w_k} \right]\,, 
\end{equation}
where the product runs over all roots $w_k$  of the entire function
$f$. The Fisher zeros, i.e., the zeros of $Z$, are easily identified
as $\beta_k = \beta+ w_k$. The Taylor coefficients $c_n^{(\beta)}$ of
the expansion of $f(w)$ in $w$, or equivalently the Taylor coefficients
of $\log Z$ in the gauge coupling variable around the simulation point
$\beta$, are nothing but the cumulants of the gauge
action up to numerical factors,
\begin{equation}
  \begin{aligned}
c^{(\beta)}_n &= \frac{(-1)^n}{n!} \llangle G^n \rrangle     \,,\\
\llangle G^n\rrangle &\equiv (-\partial_\beta)^n \log Z(\beta)\,.
  \end{aligned}
\end{equation}
Using Eq.~\eqref{eq:weier_beta} we can then relate cumulants and Fisher
zeros as follows,
\begin{equation}
\llangle G^n\rrangle = - (n-1)!\sum_k
\frac{1}{(\beta_k-\beta)^n} \quad \quad (n \geq 2)\,. 
\end{equation}
If the quark determinant is real, in particular for $\mu=0$, we have the
same symmetry we had with $\zeta$ and $\hat{\mu}^2$, i.e., Fisher zeros
also come in complex conjugate pairs, and at high enough order in the
cumulants we have: 
\begin{equation}
\label{eq:Fisher_pair}
c^{(\beta)}_n  = \frac{-1}{n} \frac{2 \cos(n \theta)}{r^n}\,,
\end{equation} 
where now for the leading Fisher zeros in the upper complex plane we have 
$\beta_k - \beta = r e^{i \theta}$ with $\theta\in [0,\pi]$. 
This is the same formula that we found above for the expansion
coefficients in our two chemical-potential-type variables, and therefore
our previous discussion about the inapplicability of the ratio
estimator also applies in this case. By the same token, our discussion
in the next subsection about how to estimate the position of the
leading Lee-Yang zero position from the cumulants of the baryon number
(i.e., the Taylor coefficients in $\hat{\mu}$) will also apply to the
estimate of the position of the leading Fisher zero from the
cumulants of the gauge action.  

The relation between high-order cumulants of the energy and the Fisher zeros
in statistical mechanics systems was also pointed out recently 
in
Refs.~\cite{PhysRevLett.110.050601,PhysRevLett.118.180601,PhysRevE.97.012115}. 

\subsection{Other estimators from the literature}

While the ratio estimator is not universally applicable, the
Cauchy-Hadamard theorem (see, e.g., Ref.~\cite{Ahlfors_analysis})
guarantees that the convergence radius of a Taylor series $\sum_n a_n
x^n$ can always be obtained as $1/r=\lim \sup_{k \to \infty}
|a_k|^{1/k}$.  
We will refer to 
\begin{equation}
\label{eq:ch}
r_k^{\rm (CH)}=\left|\frac{1}{c_k}\right|^{\frac{1}{k}}\,,  
\end{equation}
as the Cauchy-Hadamard estimator. 
Another estimator found in the literature is the Mercer-Roberts
estimator~\cite{mercer1990centre}: 
\begin{equation}
\label{eq:ormr}
r^{\rm (MR)}_k = \left| \frac{c_{k+1}c_{k-1}-c_k^2}{c_{k+2}c_k
    -c_{k+1}^2} \right|^{\frac{1}{2}}\,. 
\end{equation}
The coefficients $c_k$ are here understood to be any of
$c^{(\hat{\mu})}_k$, $c^{(\zeta)}_k$ 
or $c^{(\beta)}_k$. In the case at hand, the 
Mercer-Roberts estimators  
can be shown to have a well-defined large-order limit. 

\subsection{Exact estimators for a single  
zero}
As discussed in the previous subsections, the high-order behavior of
the Taylor expansion of the partition function, either in the fugacity
parameter or in the chemical potential, is determined by the Lee-Yang
zeros closest to the origin, and is approximately described by
Eqs.~\eqref{eq:taylor_1ly} and \eqref{eq:taylor_1ly_mu}. 
Similarly, the behavior of the high-order cumulants of the gauge 
action is determined by the leading pair of Fisher zeros,
as given by Eq.~\eqref{eq:Fisher_pair}. 
Since the leading zeros are related by the symmetries of the partition
function, 
with a slight abuse of terminology we will refer to 
Eqs.~\eqref{eq:taylor_1ly}, \eqref{eq:taylor_1ly_mu} and
\eqref{eq:Fisher_pair} as the contribution of a single zero.  
Furthermore, given the similarities of the three cases
($\zeta$,$\hat{\mu}$ and $\beta$), it is possible to give a unified
treatment. 

Knowledge of the high-order behavior of the Taylor expansion allows us
to design estimators that for the case of a single zero give
exactly the convergence radius, and receive corrections only from
singularities farther apart from the origin. In this paper we present
three such estimators. 
The first estimator is a modified version of the Mercer-Roberts
estimator, Eq.~\eqref{eq:ormr}.
Using the trigonometric identity 
\begin{equation}
\frac{\cos\left( (k+1) \theta \right) \cos \left( (k-1) \theta \right)
  - \cos^2 \left( k \theta \right) }{\cos \left((k+2) \theta \right)
  \cos \left( k \theta \right) - \cos^2 \left( (k+1) \theta \right) }
= 1 \,,
\end{equation}
and expressing $\cos(k\theta)$ in terms of the Taylor coefficients
$c_k$ of either one of Eqs.~\eqref{eq:taylor_1ly},
\eqref{eq:taylor_1ly_mu} or \eqref{eq:Fisher_pair},
one obtains the exact
estimator 
\begin{equation}
\label{eq:mmr_est}
    r^{\rm (MMR)}_k = \left| \frac{(k+1)(k-1)c_{k+1}c_{k-1}-k^2
        c_k^2}{(k+2)k c_{k+2}c_k - (k+1)^2 c_{k+1}^2} \right|^{\frac{1}{2}}\,,
\end{equation}
which equals $r^{\rm (MMR)}_k=r$ for any $k$. We will refer to this
estimator as the ``modified Mercer-Roberts estimator'' in the
following, as it is quite similar to the Mercer-Roberts estimator.

Using instead the trigonometric identity
\begin{equation}
\label{eq:2i_ide}
\cos \left( 2 k \theta \right) - 2 \cos^2 \left( k \theta \right) = -1\,,
\end{equation}
one obtains a different, but equally exact estimator:
\begin{equation}
\label{eq:2i_est}
    r^{\rm (2i)}_k = \left| \frac{2}{2k c_{2k} + k^2 c_{k}^2}
    \right|^{\frac{1}{2k}}\,. 
\end{equation}
Also for this estimator, which we will refer to as the ``doubled index  
estimator'', one has $r^{\rm (2i)}_k=r$ for any $k$. 

Finally, from the identity
\begin{equation}
\frac{\cos \left( 2 k \theta \right) - 2 \cos^2 \left( k \theta \right)}
{\cos \left( 2 (k+1) \theta \right) - 2 \cos^2 \left( (k+1) \theta
  \right)} = 1\,, 
\end{equation}
which follows trivially from Eq.~\eqref{eq:2i_ide},
one arrives at the estimator we will call the ``doubled index ratio
estimator'': 
\begin{equation}
\label{eq:2i_est_rat}
    r^{\rm (2irat)}_k = \left| \frac{2k c_{2k} + k^2 c_{k}^2}{2(k+1)
        c_{2(k+1)} + (k+1)^2 c_{k+1}^2} 
    \right|^{\frac{1}{2}}\,. 
\end{equation}

As already mentioned above, in the formulas Eqs.~\eqref{eq:mmr_est},
\eqref{eq:2i_est} and \eqref{eq:2i_est_rat} one can use for the $c_k$
any of the coefficients  $c^{(\hat{\mu})}_k$, 
$c^{(\zeta)}_k$ or  
$c^{(\beta)}_k$ to estimate the distance from the origin of the
leading zero in the variables $\hat{\mu}^2=(\mu/T)^2$,
$\zeta=e^{\hat{\mu}}-1$, or $w=\beta_C-\beta$, respectively. 

All of these three estimators give exactly the radius of convergence 
in a finite volume at any order $k$ only in the case of a single
zero. Their expressions in the general case are easily
obtained by summing over all the zeros of the partition
function (as a function of the appropriate variable). The Taylor
coefficients of the expansion in either $\zeta$, $\hat{\mu}^2$ 
or $w$ become in this case\footnote{Here
  it is 
understood that only the logarithmic terms in Eqs.~\eqref{eq:sum_fug}
and \eqref{eq:sum_mu} are considered.} 
\begin{equation}
  \label{eq:all_zeros}
  C_k^{(X)} = \sum_j \frac{-2}{k}
    \frac{\cos(k \theta_j)}{r_j^k}\,, \qquad X= \zeta,\,\hat{\mu},\,w,
\end{equation}
where the sum extends over all the zeros $A_j = r_j
e^{i\theta_j}$ in the upper complex half-plane. 
The estimators of Eqs.~\eqref{eq:mmr_est}, \eqref{eq:2i_est} and
\eqref{eq:2i_est_rat} in the general case are
simply obtained by replacing $c_k^{(X)}\to C_k^{(X)}$, and should become very
stable and accurate as soon as the effect of the subleading 
zeros becomes negligible.

With all the zeros included, the three estimators have finite-order
corrections that most likely differ among them. It is then worth
checking how and how fast the three estimators approach their
asymptotic value. 
In a realistic case when only a few low-order Taylor coefficients are
known, agreement between the three estimators would be a good reliability
check, and an indication that they are already dominated by a single
zero.
 
\subsection{Estimators for the phase  
of the closest 
zero}

Although a finite radius of convergence tells us about the presence
of a singularity at a finite distance from the expansion point,
it cannot by
itself guarantee the existence of a critical point on the real
positive $\hat{\mu}$ or $\beta$ axis. It is then worth estimating also
the phase of the leading zero, to see how far it is from the real
axis of the relevant expansion variable. 
Using the trigonometric formula
\begin{equation}
\frac{\cos \left( (k+1)\theta \right)+\cos \left( (k-1)\theta
  \right)}{2\cos \left( k \theta \right)}  = \cos\theta\,,
\end{equation}
one finds the following estimator for $\cos\theta$,
\begin{equation}
\label{eq:cosest}
\Gamma_k =    \frac{(k+1) c_{k+1} r^2 + (k-1)c_{k-1}}{2kc_k r}
\mathop\to_{k\to\infty}\cos\theta \,,
\end{equation}
where $r$ can be any of the previous estimators for the convergence
radius, and the $c_k$ can be either 
$c^{(\zeta)}_k$, $c^{(\hat{\mu})}_k$ or $c^{(\beta)}_k$. 
The cosine of the phase  
gives us all the useful
information: in fact, in the infinite-volume limit either the leading 
zero tends to the real axis, so that $\cos\theta\to \pm 1$
and we can tell where the singularity is, or it does not, in which
case the reality of the partition function requires the presence of
singularities both at $\theta$ and $-\theta$. 

Of course, in order for the convergence radius in
$\zeta=e^{\hat{\mu}}-1$ to be limited by an actual critical point,
the phase $\theta$ of $\Lambda_1$ must tend to $\pi$ in the
infinite-volume limit.  
In the case of the chemical potential $\hat{\mu}$, the formula
Eq.~\eqref{eq:cosest} gives the cosine of the phase of the leading
zero in $\hat{\mu}^2$, i.e., $\cos 2\theta$ with $\theta$ the phase of
$\lambda_1$.  In this case $\cos 2\theta\to 1$ in the infinite-volume
limit signals the presence of a genuine phase transition at real
chemical potential, while $\cos 2\theta\to -1$ signals the presence of
a genuine phase transition at imaginary chemical potential. 

Since the convergence radius for $\hat{\mu}$ and $\zeta$
may not be determined by the same singularity, one could in principle
extrapolate to infinite volume the  
phase of the leading Lee-Yang zero found with either variable, to
check if any of the two Taylor series is limited by a true  
critical point.

The  
phase estimator Eq.~\eqref{eq:cosest} can of course also be used to 
estimate the phase  
of the leading Fisher zero  
in the complex gauge coupling variable $w=\beta_C-\beta$, 
giving also the explicit location of the leading complex zero of
$Z(\beta)$. 

\subsection{Statistical errors at high orders}
Knowing the asymptotic behavior of the Taylor coefficients,
Eqs.~\eqref{eq:taylor_1ly}, \eqref{eq:taylor_1ly_mu} and
\eqref{eq:Fisher_pair}, one can also estimate the statistical error
for asymptotically large orders, assuming linear error
propagation. Assuming we can determine the 
polar coordinates of the leading 
zero $(r,\theta)$ with
covariance matrix $\left( \begin{smallmatrix} \sigma^2_r & \sigma^2_{r
      \theta} \\ \sigma^2_{r \theta} &
    \sigma^2_{\theta} \end{smallmatrix}\right)$, for the case of the
expansion coefficients $c_k$ this leads to
\begin{equation}
\frac{\sigma^2_{c_k}}{c_k^2} \approx k^2 \left( \tan^2 (k \theta)
  \sigma^2_\theta + \frac{\sigma^2_r}{r^2} + 2 \frac{\sin (k \theta)
    \cos(k \theta) \sigma_{r \theta}^2}{r}\right) \,.
\end{equation}
This means that for a fixed volume and a fixed ensemble,
asymptotically high orders of the expansion coefficients should have a
relative statistical error of $O(k)$ (with some oscillations around and above
the linear behavior, especially because of the $\tan(k \theta)$ term,
which can be large). On the other hand our estimators will have, by 
the same linear error propagation assumption,
\begin{equation}
\sigma^2_{r^{\rm{(MMR)}}_k} \approx \sigma^2_{r^{\rm{(2i)}}_k} \approx
\sigma^2_r \,,
\end{equation}
at sufficiently large $k$. In other words, 
in contrast to the coefficients $c_k$ themselves which on 
a given ensemble have
larger and larger error bars for larger $k$,
the statistical errors of our convergence radius estimators at 
asymptotically large $k$ will converge to some finite number, namely 
the error of the distance from the expansion point of the closest
zero. This is due to the  
large correlations between 
the statistical errors of the $c_k$ 
for different orders $k$.

\section{Convergence radius for $N_f=4$ unimproved staggered fermions  on $N_t=4$ lattices} 
\label{sec:unimp}

To illustrate the issues raised in the general section, we now perform
an analysis of the proposed estimators on a toy lattice model, where we
can analyze fluctuations to arbitrarily high order.

\begin{figure}
  \centering
      \includegraphics[width=0.5\textwidth]{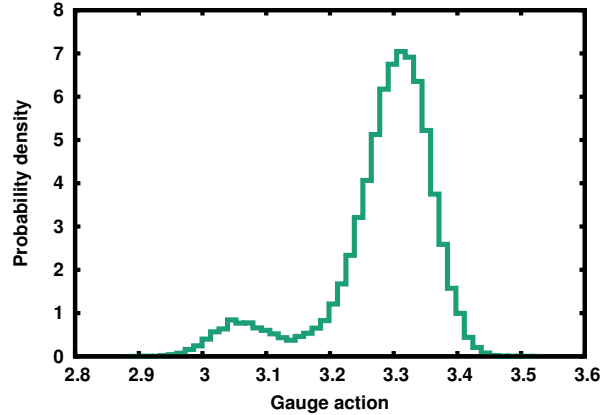}
  \caption{Histogram of the gauge action in our simulation at
    $\beta=4.940$, indicating we are close to  
	         but not quite at $\beta_c$.}
\label{fig:gact}
\end{figure}

\subsection{Choice of the toy model}
The model we used is $N_f=4$ unimproved staggered fermions
on lattices with a fixed number $N_t=4$ of time slices,
in a small volume $V=6^3$, with a bare fermion mass of $ma=0.01$. 
According to the literature~\cite{Kogut:1987rz}, for this choice of
mass this model is expected to have a line of first-order phase transitions starting from $\mu=0$, and therefore the
leading Lee-Yang zero should be rather close to the real axis  
even in a small volume. 
Since we are not performing any rooting,
the partition function is indeed of the form Eq.~\eqref{eq:part_func}. This would also be the case for 
Wilson fermions with arbitrary $N_f$.

We chose a value of the gauge coupling $\beta=4.940$ that is slightly
below the transition at $\mu=0$ (see Fig.~\ref{fig:gact}), so that the
real part of the leading Lee-Yang zeros should converge to a nonzero
value in the large volume limit. We use an ensemble of 16000 configurations, 
each separated by 10 HMC trajectories.

\begin{figure}[t]
  \centering
  \includegraphics[width=0.5\textwidth]{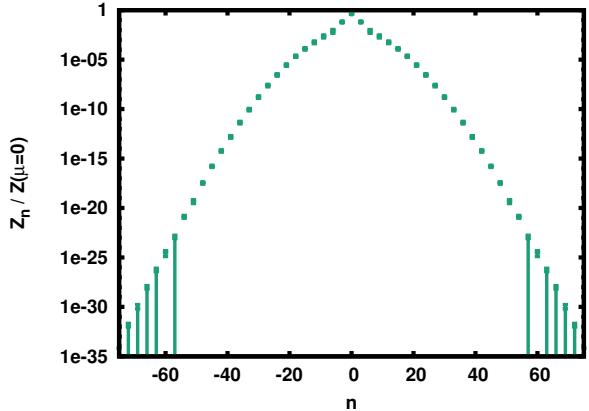}
  \caption{
 Canonical partition functions, normalized to $Z(\hat{\mu}=0)$. Further
 coefficients have error bars over $100\%$, signalling the onset of
 the sign problem.
  }
    \label{fig:coeffs}
\end{figure}

\subsection{Calculation of the cumulants of the quark number to arbitrary order} 
For simplicity, we will consider the case of four flavors of quarks with
degenerate  
quark chemical potentials $\mu_1=\mu_2=\mu_3=\mu_4\equiv\mu$. 
If we choose the same quantum numbers as in the 
1+1+1+1 flavor physical case, i.e., if
\begin{equation}
\begin{aligned}
 \mu_d &= \frac{1}{3} \mu_B - \frac{1}{3} \mu_Q \,,\\
 \mu_u &= \frac{1}{3} \mu_B + \frac{2}{3} \mu_Q \,,\\
 \mu_s &= \frac{1}{3} \mu_B - \frac{1}{3} \mu_Q - \mu_S \,,\\
 \mu_c &= \frac{1}{3} \mu_B + \frac{2}{3} \mu_Q - \mu_C\,,
\end{aligned}
\end{equation}
then this choice of chemical potentials corresponds to $\mu_B=3\mu$,
with $\mu_Q=\mu_S=\mu_C=0$. 

Our calculation is based on the reduction formula
of Ref.~\cite{Hasenfratz:1991ax}, which 
reads\footnote{The
  formula appearing in Ref.~\cite{Hasenfratz:1991ax} misses the factor
of 3 in the exponential before the product.}
\begin{equation}
\det M(\hat{\mu}) = e^{-3 V \hat{\mu} } \prod_{i=1}^{6 V} \left(
  \xi_i - e^{\hat{\mu}} \right) \,,
\end{equation}
where $V= N_s^3$ with $N_s$ the spatial linear size of the lattice,
and where the $\xi_i$ are eigenvalues of the reduced matrix $P$
given below. 
Since $P$ is $\mu$-independent, once the $\xi_i$ are known
for a given gauge configuration, one can calculate the corresponding 
unrooted quark determinant for arbitrary $\mu$, which allows us to
extract fluctuations to arbitrarily high order on the given ensemble.
Working in the temporal gauge [$U_4(t,\vec{x})=1$ for $0\le t < N_t$], 
$P$ is obtained as (spatial indices are suppressed) 
\begin{equation}
P = -\prod_{i=0}^{N_t-1}
 \begin{pmatrix} B_i & 1 \\ 1 &
    0 \end{pmatrix} L \,,
\end{equation}
with $B_i$ the sum of the spatial derivatives and mass parts of the 
staggered matrix on the $i$-th time-slice, and $L$ the temporal link on
the last time slice (i.e., the untraced Polyakov loop). Performing a
Monte Carlo simulation at $\mu=0$, one can then obtain the partition
function at nonzero $\mu$ via
\begin{equation}
  \label{eq:mupartfunc}
  \begin{aligned}
  \frac{Z(\hat{\mu})}{Z(0)} &= \left\langle \frac{\det M(\hat{\mu})
    }{\det M(0)}\right\rangle_0 
  = e^{-3 V \hat{\mu} } \left\langle  \prod_{i=1}^{6 V}\frac{
  \xi_i - e^{\hat{\mu}}}{ \xi_i - 1 }\right\rangle_0 \\
&=   e^{-3 V\hat{\mu} }
\,{\cal P}(e^{\hat{\mu}})
\,,
  \end{aligned}
\end{equation}
where the subscript 0 indicates that the expectation value is computed
at $\mu=0$. It is evident that ${\cal P}$ is a polynomial in the
fugacity,
whose coefficients are the (normalized) canonical partition functions
$Z_n/Z(0)$ [see Eq.~\eqref{eq:part_func}]. 
While here we obtain them by averaging appropriate combinations of the
eigenvalues of $P$, they are usually obtained by Fourier transform
from simulations at imaginary chemical
potential~\cite{Hasenfratz:1991ax,Morita:2012kt,Fukuda:2015mva,Nakamura:2015jra,deForcrand:2006ec,Ejiri:2008xt,Li:2010qf,Li:2011ee,Danzer:2012vw,Gattringer:2014hra,Boyda:2017lps,Goy:2016egl,Bornyakov:2016wld,Boyda:2017dyo,Wakayama:2018wkc,Nagata:2014fra}. 
Because of the Roberge-Weiss symmetry, all coefficients with
$n$ not a multiple of $3$ vanish upon gauge averaging and 
can thus be set to zero. 
The normalized canonical partition functions are shown in
Fig.~\ref{fig:coeffs}. 

\begin{figure}[t]
  \centering
  \includegraphics[width=0.5\textwidth]{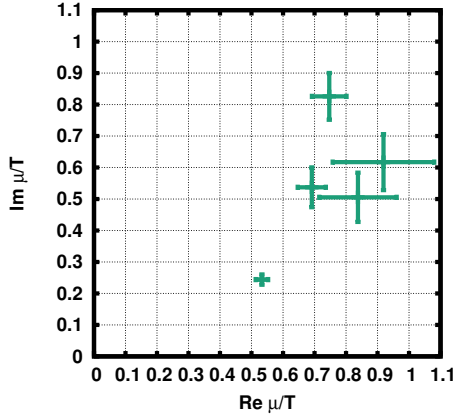}
  \caption{The first five Lee-Yang zeros in the chemical potential
    $\hat{\mu}=\mu/T$ with their statistical errors.
    \label{fig:LY0s}
					}
\end{figure}

The radius of convergence of a Taylor expansion of $\log {\cal P}$ is
determined by the position of the zeros of ${\cal P}$, and it is
clearly the same as that of $\log Z$, so that we can  simply
ignore the exponential prefactor in Eq.~\eqref{eq:mupartfunc}. 
On such a small lattice, it was possible to find all roots of $\mathcal{P}$ 
using standard methods.
The roots can then be used, e.g., in Eq.~\eqref{eq:all_zeros} to obtain the
Taylor coefficients of the expansion in the fugacity parameter $\zeta$ 
to arbitrarily high order.  
The Lee-Yang zeros closest to $\mu=0$ can be seen in
Fig.~\ref{fig:LY0s}.  

\begin{figure}[t]
  \centering
  \includegraphics[width=0.5\textwidth]{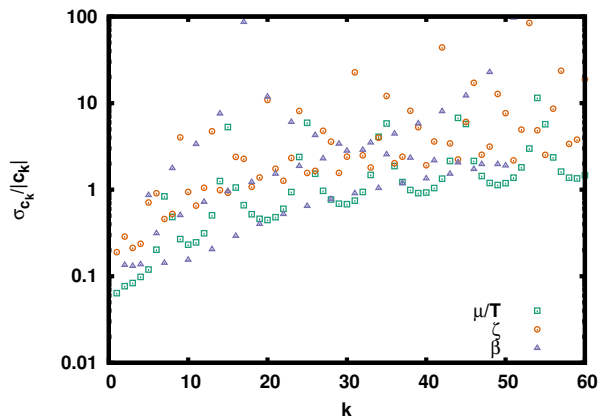}
  \caption{Relative statistical error of the Taylor coefficients
    $c^{(\hat{\mu})}_k$, $c^{(\zeta)}_k$ and $c^{(\beta)}_k$ as a
    function of the order of the expansion. Notice that they climb to
     well over $100\%$. 
  }
  \label{fig:rel_err} 
\end{figure}

\subsection{Calculation of the cumulants of the gauge action to arbitrary order} 
The calculation of the coefficients $c^{(\beta)}_k$ is more straightforward. After
a direct computation of the moments $\langle G^n \rangle$ one can use
the following recursive formula~\cite{smith1995recursive},
\begin{equation}
\label{eq:cumulant_recursion}
\llangle G^n \rrangle = \langle G^n \rangle - \sum_{m=1}^{n-1}
\binom{n-1}{m-1} \llangle G^m \rrangle \langle G^{n-m} \rangle \,,
\end{equation}
to build all cumulants. 
Arbitrary precision arithmetic has to be used both in the calculation
of the moments $\langle G^n \rangle$ and in applying the recursion
formula Eq.~\eqref{eq:cumulant_recursion}. This might be somewhat
counter-intuitive, as much more digits have to be kept in the
calculation than the statistical error bars of the cumulants would 
suggest. On the other hand, the errors of the cumulants of different
order are very strongly correlated, and this correlation should not be
altered in order to get the radius of convergence with a reliable
error estimate.

\subsection{Numerical results for the convergence radius estimators}
We come now to the numerical results on the convergence radius
estimators mentioned in Section 2. First, we show in
Fig.~\ref{fig:rel_err} the relative statistical error of the Taylor
coefficients $c^{(\zeta)}_k$, $c^{(\hat{\mu})}_k$ and
$c^{(\beta)}_k$. As expected, the statistical errors increase  
with the order of the expansion, getting to order $O(100\%)$ and above
quite soon. In spite of this, as we will show below, it is possible to
take combinations of them (i.e., our radius estimators) that have
small error bars and give the radius of convergence directly.

\begin{figure}[t]
  \centering
  \includegraphics[width=0.5\textwidth]{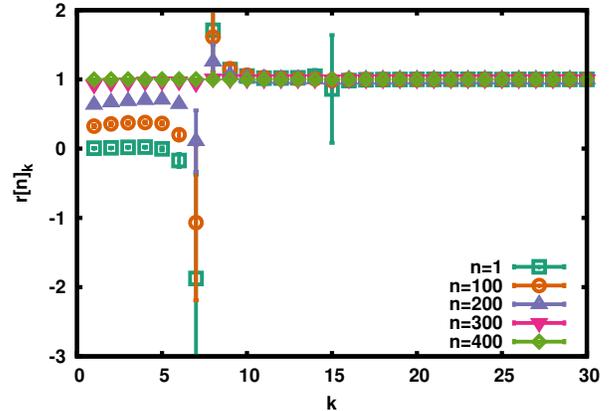}
  \includegraphics[width=0.5\textwidth]{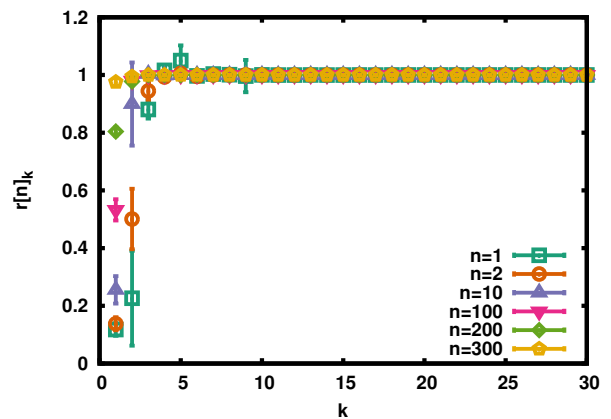}
    \caption{The ratios $r^{(X)}[n]_k$ for several values of $n$ and $k$
    in case of $X=\zeta$ (top) and
    $X=\hat{\mu}$ (bottom).\label{fig:contrib}}  
\end{figure}

\begin{figure}[t]
  \centering
  \includegraphics[width=0.5\textwidth]{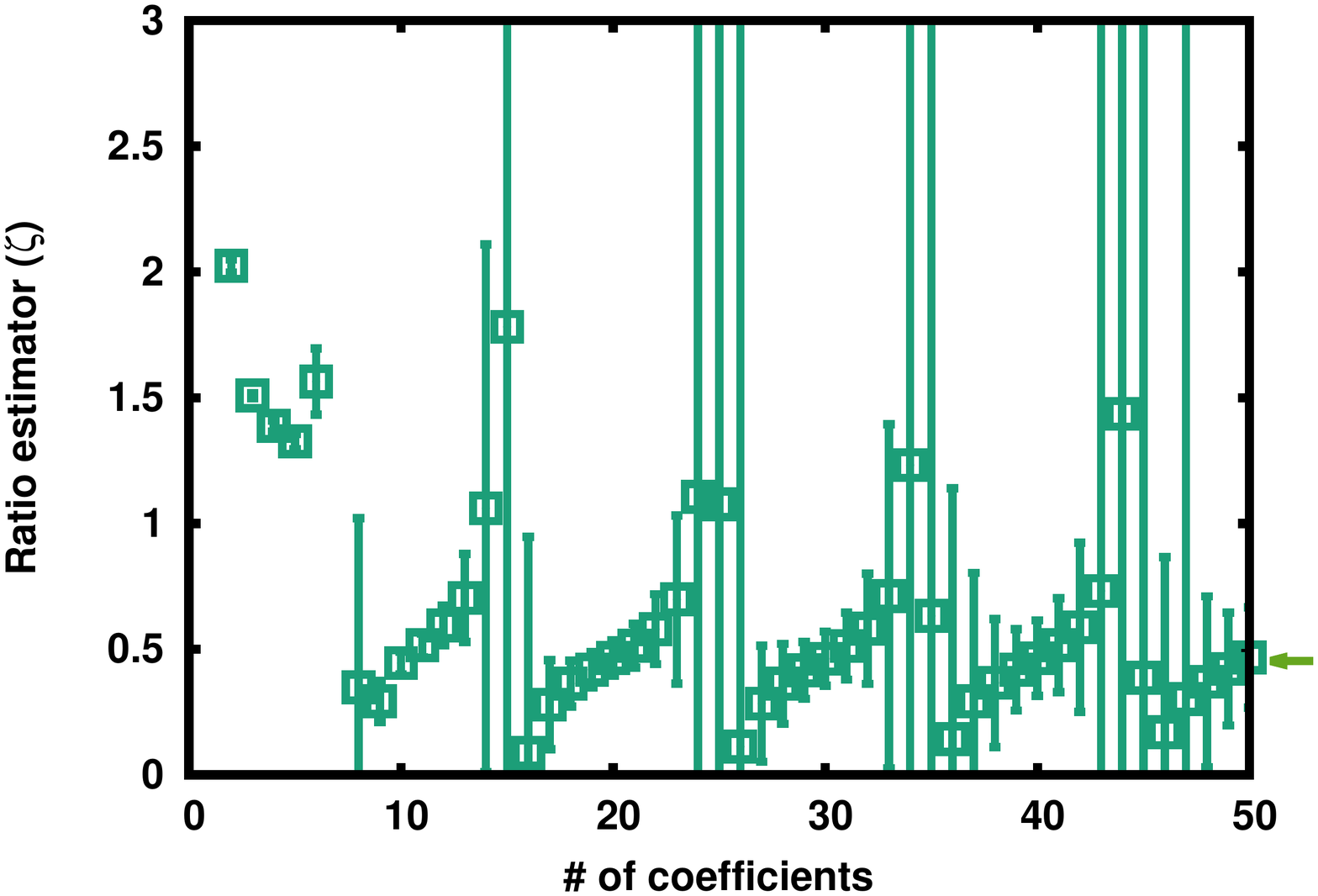}
  \includegraphics[width=0.5\textwidth]{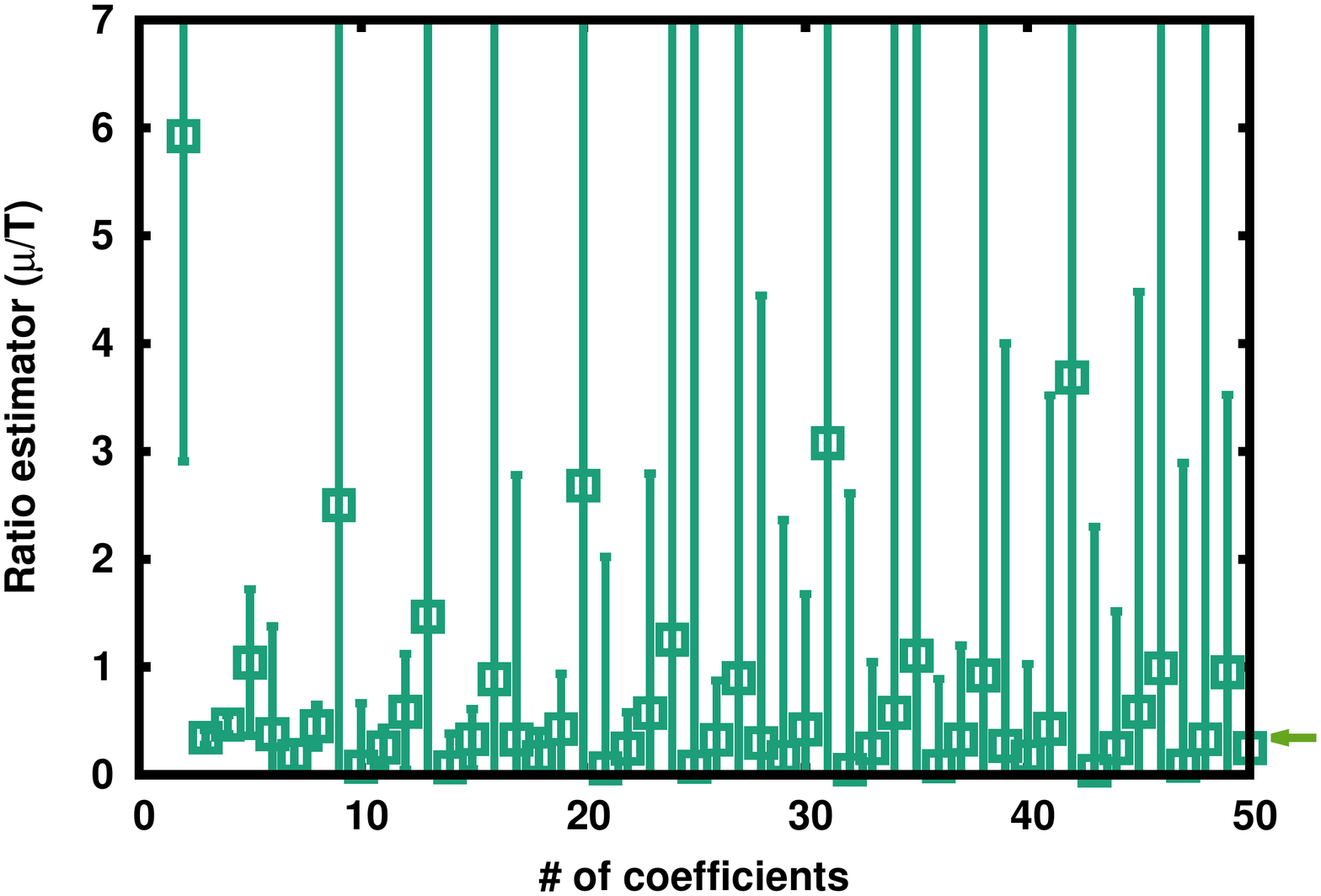}
  \caption{Simple ratio estimators compared to the correct radius of
    convergence, as given by the analysis of the Lee-Yang zeros (the light green
    arrow) for the case of the fugacity parameter
    $\zeta=e^{\hat{\mu}}-1$ (top) and the chemical potential
    $\hat{\mu}$ (bottom). \label{fig:ratio}} 
\end{figure}

It is interesting to study how much of a particular expansion
coefficient comes from the leading $n$ Lee-Yang zeros. We quantify
this by the ratio
\begin{equation}
r^{(X)}[n]_k =  
\frac{1}{C_k^{(X)}}
\sum_{j=1}^{n} \frac{-2}{k}
    \frac{\cos(k \theta_j)}{r_j^k}\,,
\end{equation}
where $X$ can be either $\hat{\mu}$ or $\zeta$, the $C_k^{(X)}$ are
given in Eq.~\eqref{eq:all_zeros}, and the sum from $j=1$
to $n$ contains the $n$ Lee-Yang zeros in $\hat{\mu}^2$ or $\zeta$
closest to $\hat{\mu}=0$ or $\zeta=0$, limited to the upper complex
half-plane. 
These ratios for several values of $n$ can be seen in
Fig.~\ref{fig:contrib}. 

In the case of the fugacity parameter expansion, hundreds of Lee-Yang
zeros give an important contribution to low-order coefficients, before
a single Lee-Yang zero starts to dominate around $k=O(10)$. For the
chemical potential expansion the situation is much better, with only
$O(10)$ Lee-Yang zeros contributing to the second coefficient, and
the sum over zeros 
saturating sooner.
This suggests that our
estimators will work better with the chemical potential, as we shall
see. 

We now turn to the study of the radius of convergence. 
We first look at the ratio
estimators, that can be seen in Fig.~\ref{fig:ratio}. 
Notice that low orders of the ratio estimator overestimate 
the real radius of convergence, making them very misleading. 
At higher orders, notice that  the ratio estimator does not converge
to a specific 
value, just as expected from our analytical discussion in Section
\ref{sec:taylor}. Interestingly, the statistical error of the ratio
estimator does get smaller  when it is close to the correct value,
i.e., when $\left| \frac{\cos (k \theta)}{\cos ( (k+1) \theta)}\right|
\approx 1$.  
Nonetheless, it is still quite noisy there and never reaches the high
statistical  
precision of our newly proposed estimators (see below), and based on the
ratio estimator alone it would not be possible to conclude what the
radius of convergence is. 

In previous
studies~\cite{Huovinen:2009yb,Borsanyi:2011sw,Alba:2017mqu,Huovinen:2017ogf,Vovchenko:2017xad}
it was customary to compare the ratio estimators for the  
Taylor expansion in the baryon number chemical potential to the hadron 
resonance gas~\cite{Dashen:1969ep,Venugopalan:1992hy}. 
As there is no finite $\mu$ transition in this model, when such a
comparison yields results consistent with the HRG one concludes that
there are no signs of criticality in the fluctuations under
scrutiny. In the case presented here, the HRG prediction 
for the first ratio is
$\frac{c^{(\hat{\mu})}_1}{c^{(\hat{\mu})}_2}=\frac{1/2!}{1/4!}=6$, 
independently of the hadron spectrum. This matches our numerical
results within error bars. The second ratio in the HRG  
is
$\frac{c^{(\hat{\mu})}_2}{c^{(\hat{\mu})}_3}=\frac{1/4!}{1/6!}=30$. This
is two orders of magnitude 
higher than our numerical data, so one clearly sees some sign of
non-hadronic matter, but  
without our improved estimators it is unclear how one can make more
quantitative statements. 

\begin{figure}[t]
  \centering
  \includegraphics[width=0.5\textwidth]{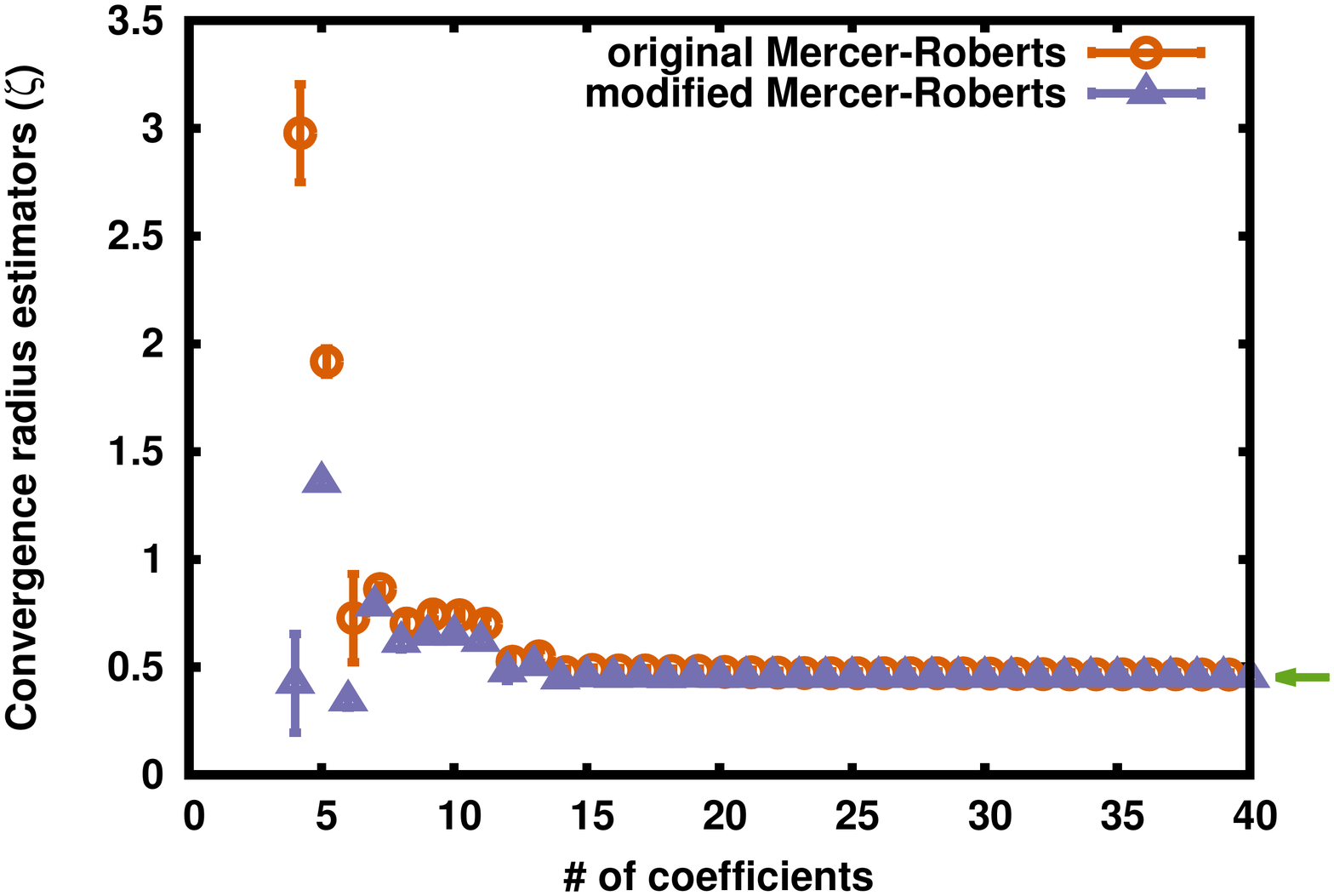}
  \includegraphics[width=0.5\textwidth]{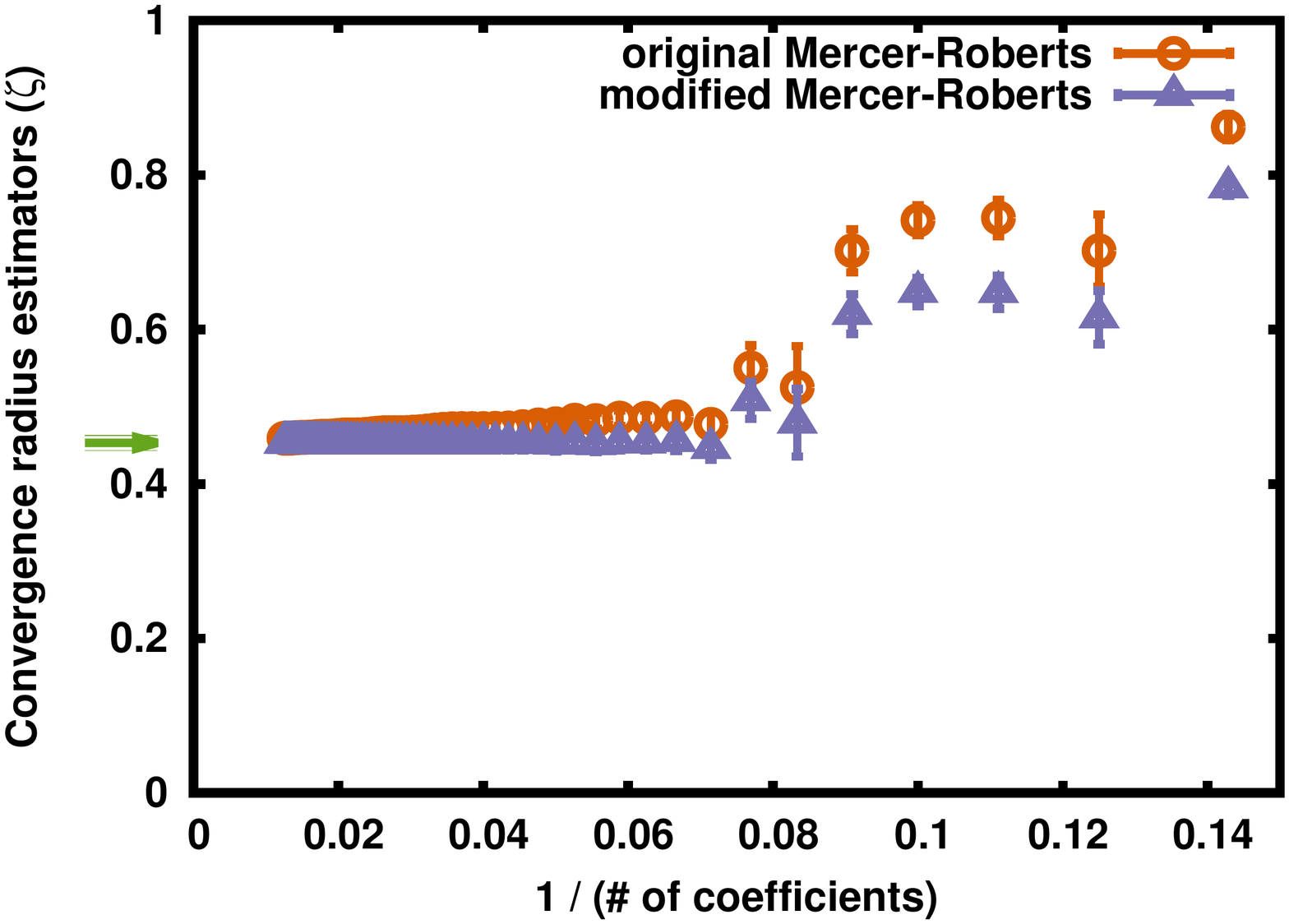}
  \caption{Top: The original and our modified Mercer-Roberts
    estimators for the convergence radius as a function 
    of the expansion order $n$ needed to calculate the given estimator
    for the variable $\zeta=e^{\hat{\mu}}-1$. Data points are slightly
    shifted for visual clarity. The estimators are compared to the correct radius of
    convergence, as given by the analysis of the Lee-Yang zeros (the light green
    arrow).
    Bottom: The same quantities, but as a function of $1/n$ for higher orders of
    the expansion.}
  \label{fig:MercerRoberts}
\end{figure}

\begin{figure}[t]
  \centering
  \includegraphics[width=0.5\textwidth]{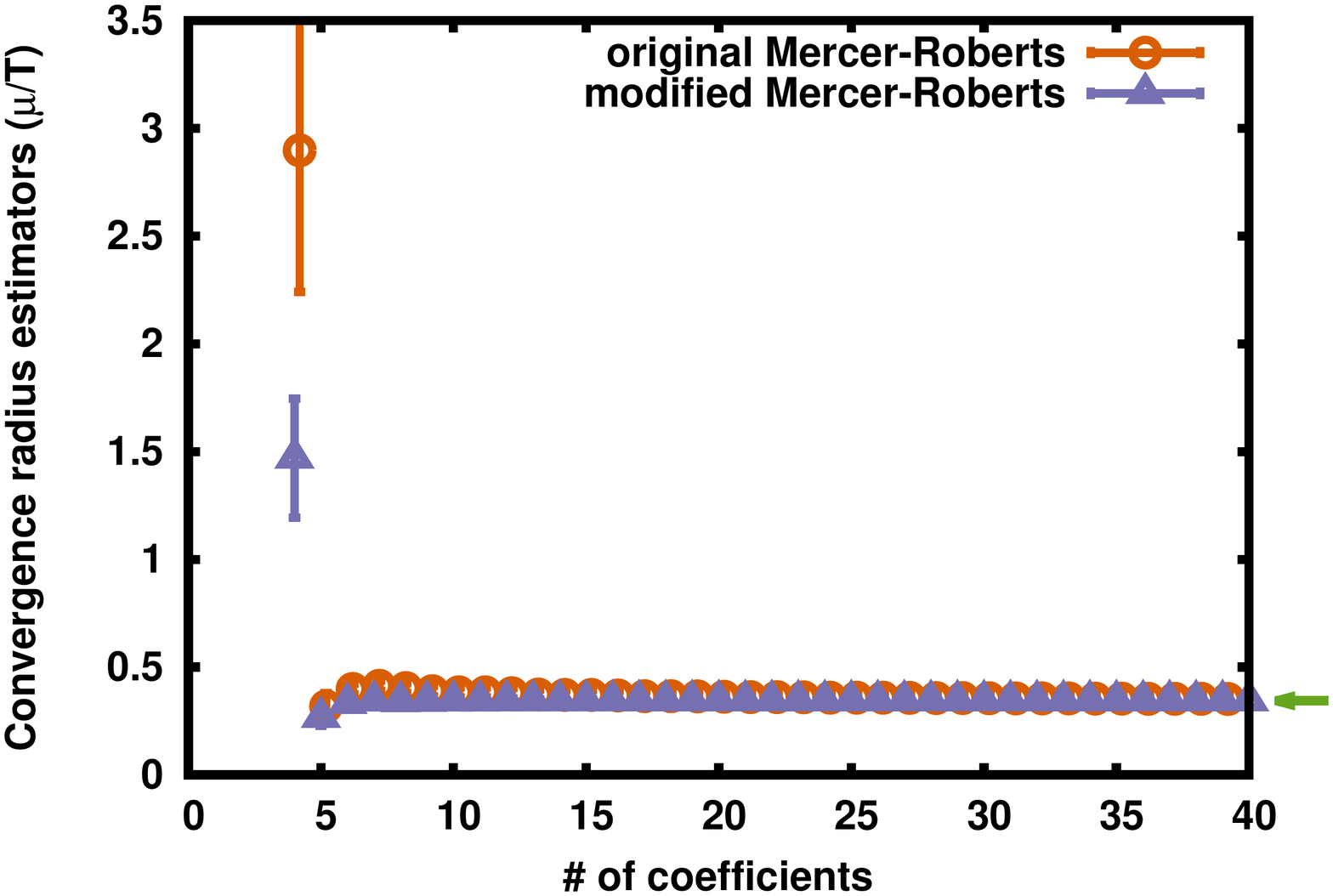}
  \includegraphics[width=0.5\textwidth]{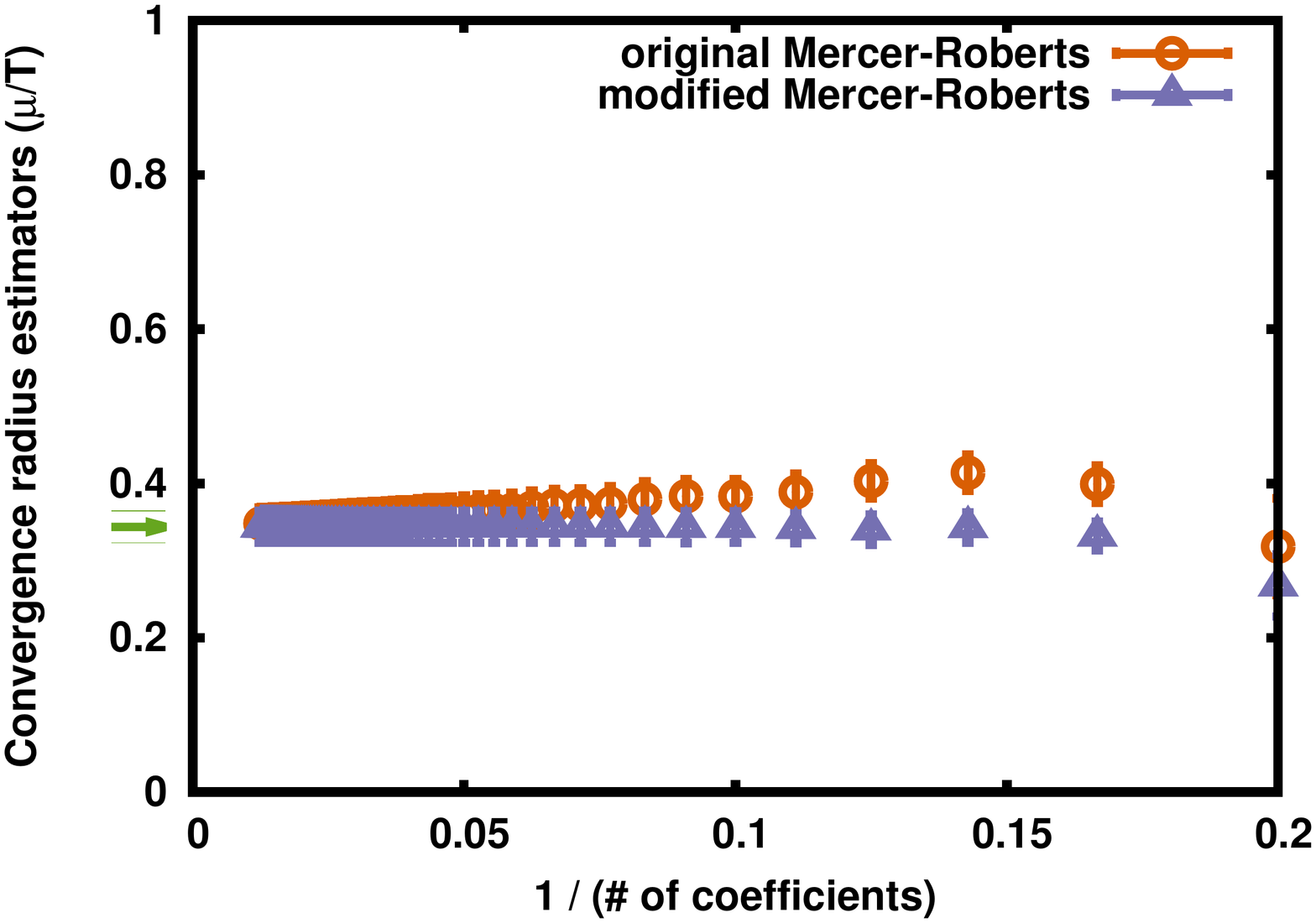}
  \caption{Top: The original and our modified Mercer-Roberts
    estimators for the convergence radius as a function 
    of the expansion order $n$ needed to calculate the given estimator
    for the variable $\hat{\mu}$. Data points are slightly
    shifted for visual clarity. The estimators are compared to the correct radius of
    convergence, as given by the analysis of the Lee-Yang zeros (the light green
    arrow).
    Bottom: The same quantities, but as a function of $1/n$ for higher orders of
    the expansion.
		}
  \label{fig:MercerRoberts_mu}
\end{figure}

We now go on to the improved convergence radius estimators proposed in
this paper. Both the original and our modified version of the
Mercer-Roberts estimator can be seen for the fugacity parameter $\zeta$ in
Fig.~\ref{fig:MercerRoberts}. As can be seen, by the 
time the original Mercer-Roberts estimator starts to become linear in
$1/n$, so that a linear fit could be performed to obtain the
convergence radius, our modified estimator already gives the correct
answer. Another way to quantify the improvement achieved with our
modification is to say that to get the correct convergence radius
within 1$\sigma$ of the statistical error bars, our estimator needs 13
orders of the expansion, while the original Mercer-Roberts needs 20.
A similar comparison for the case of the chemical potential can be seen
in Fig.~\ref{fig:MercerRoberts_mu}. We see that the convergence radius
estimators for the chemical potential work significantly better compared
to the fugacity parameter, converging to the correct value already with
a 6th order (i.e., $\hat{\mu}^{12}$) expansion. Both of the Taylor series 
have a convergence radius determined by the leading Lee-Yang zero at 
$\rm{Re}\, \hat{\mu} = 0.533 \pm 0.020$ and $\rm{Im}\, \hat{\mu} = 0.244 \pm
0.016$ (Fig.~\ref{fig:LY0s}).  

\begin{figure}
  \centering
  \includegraphics[width=0.5\textwidth]{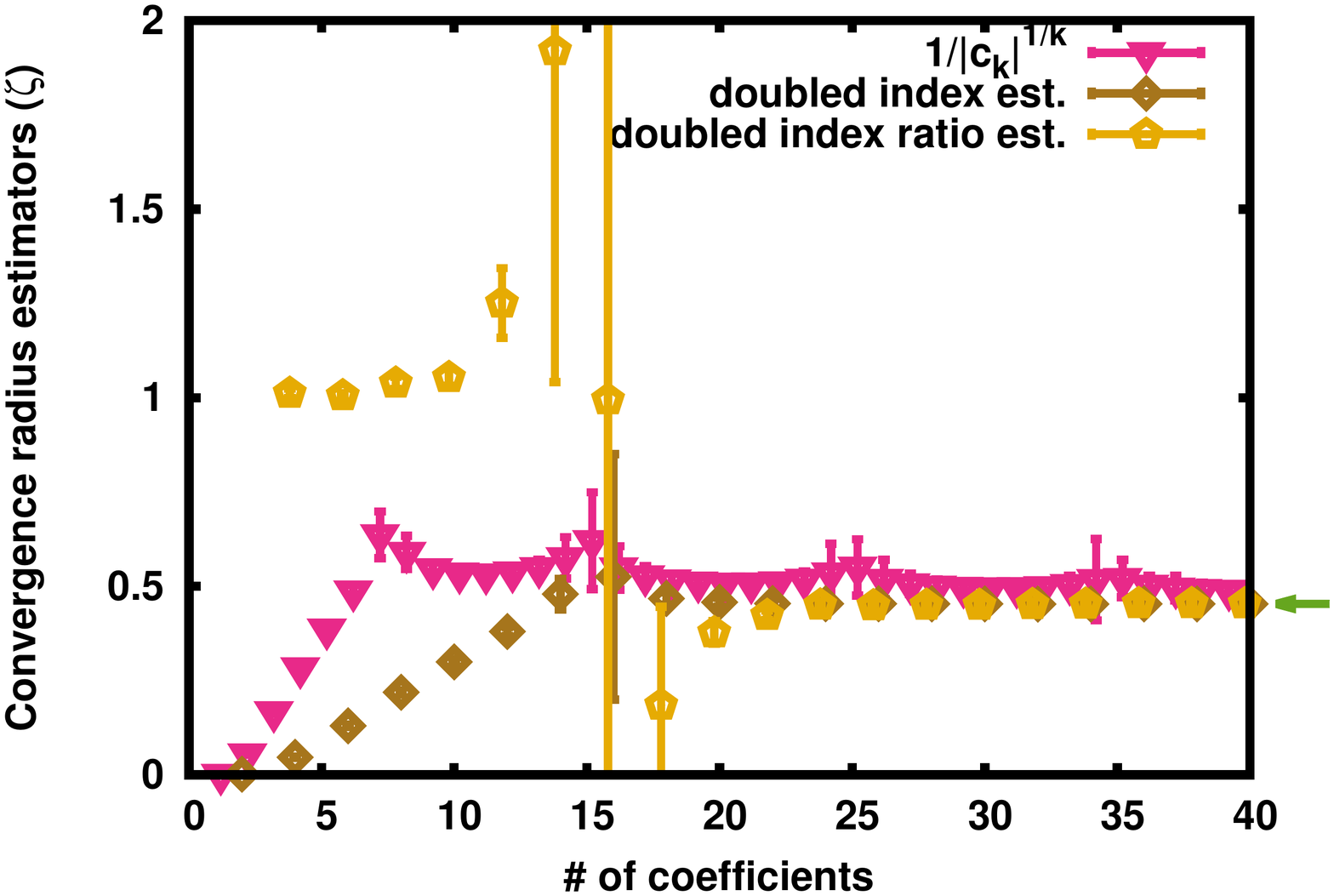}
  \includegraphics[width=0.5\textwidth]{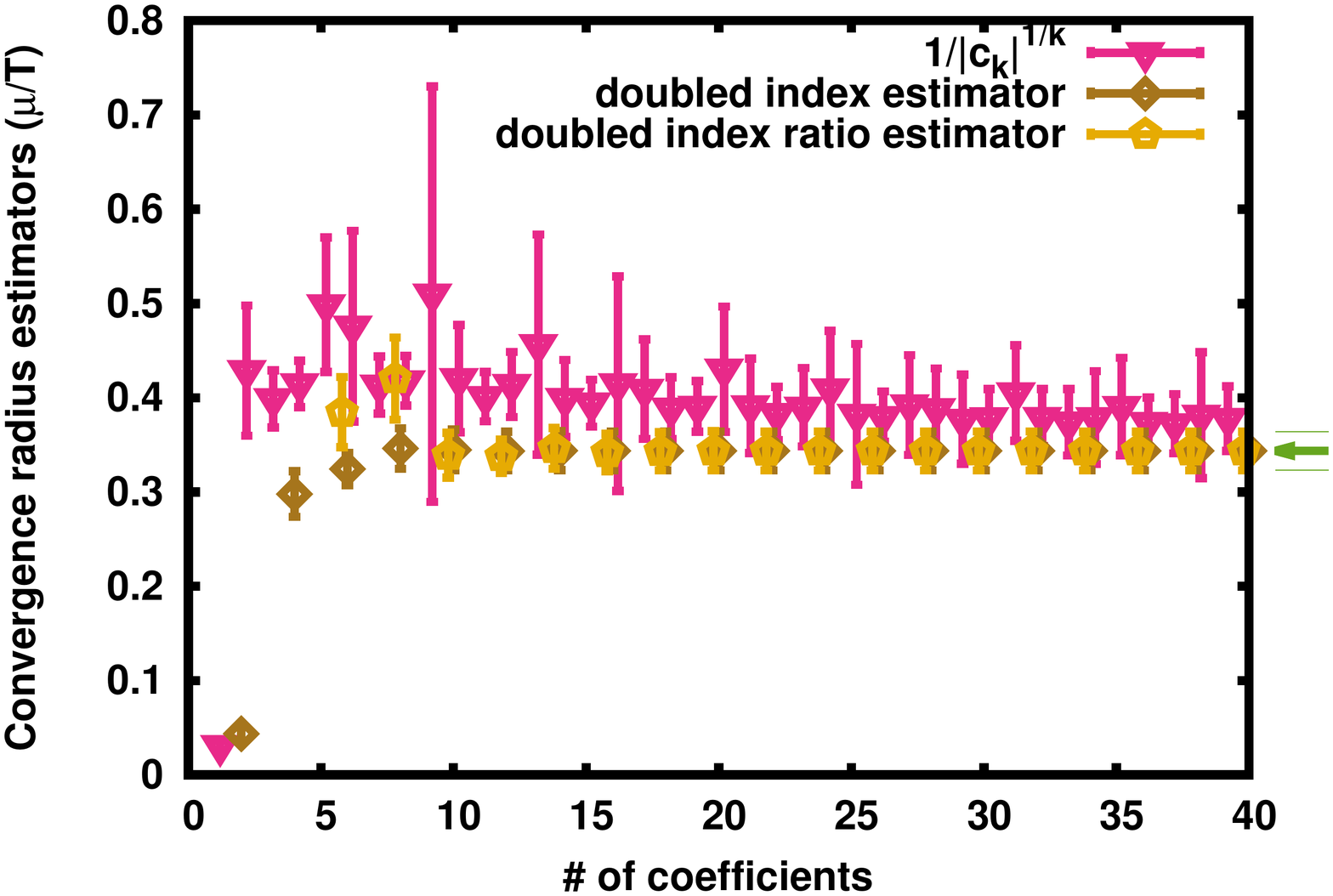}
  \caption{The Cauchy-Hadamard estimator  
    and our doubled index and doubled index ratio estimators for
    $\zeta$ (top) and $\hat{\mu}$ (bottom). Data points are slightly
    shifted for visual clarity. The estimators are compared to the correct radius of
    convergence, as given by the analysis of the Lee-Yang zeros (the light green
    arrow).
		\label{fig:2k}
		} 
\end{figure}

The doubled index estimator and the doubled index ratio estimator 
are compared to the Cauchy-Hadamard estimator for the 
fugacity parameter   
in Fig.~\ref{fig:2k} (top) and for the chemical potential in
Fig.~\ref{fig:2k} (bottom). The doubled index estimator
and the Cauchy-Hadamard estimator
behave quite similarly qualitatively, 
but our proposal is a big improvement over the other,
needing much fewer Taylor
coefficients to approach the correct value within $1 \sigma$. 

\begin{figure}[t]
  \centering
  \includegraphics[width=0.5\textwidth]{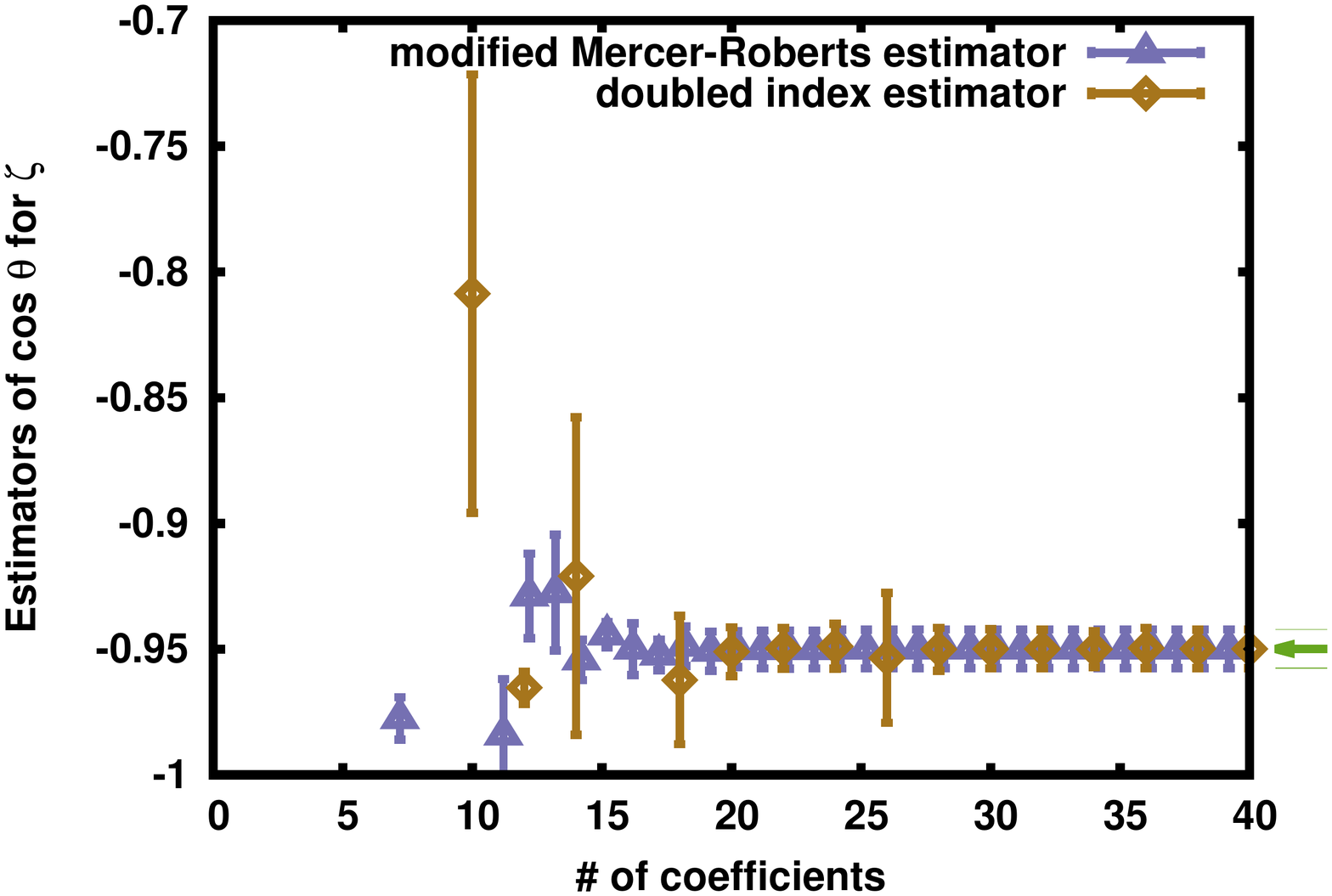}
  \includegraphics[width=0.5\textwidth]{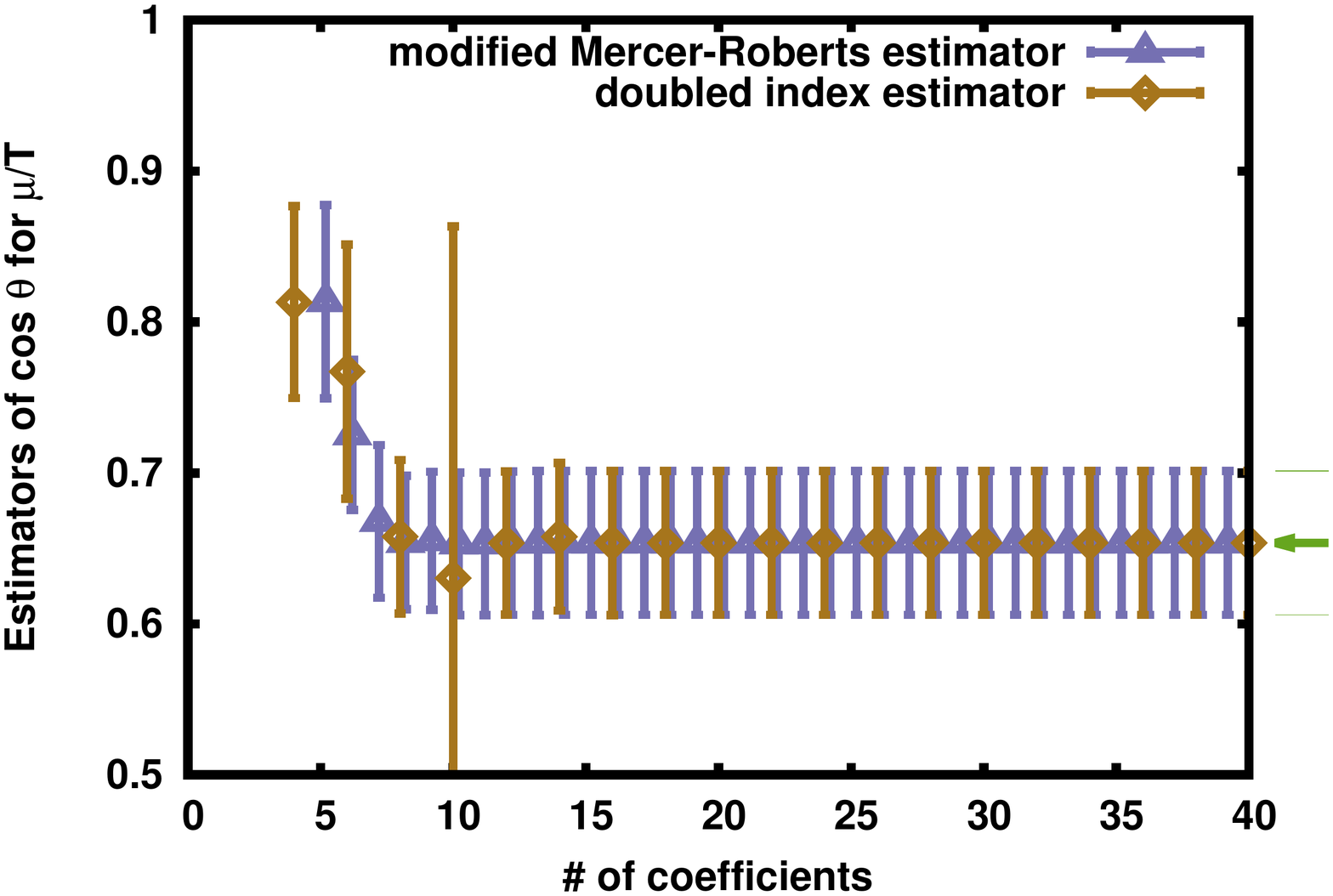}
  \caption{Estimators of $\cos \theta$ built from the doubled index
    and modified Mercer-Roberts estimators for $\zeta$ (top) and
    $\hat{\mu}$ (bottom). Data points are slightly
    shifted for visual clarity. The estimators are compared to the correct radius of
    convergence, as given by the analysis of the Lee-Yang zeros (the light green
    arrow).
    For low orders (not shown in the plot) certain estimators give values
    outside the plot range. } 
  \label{fig:cos}
\end{figure}

Two estimators for 
$\cos \theta$ can be seen in Fig.~\ref{fig:cos}.
For low orders in the Taylor series these estimators happen to be outside
the range $[-1:1]$, which very clearly indicates that the leading
Lee-Yang zeros is not dominating the series, as only in the case of  
a single Lee-Yang zero will these combinations reduce to a single
cosine. 

\begin{figure}[t]
  \centering
  \includegraphics[width=0.5\textwidth]{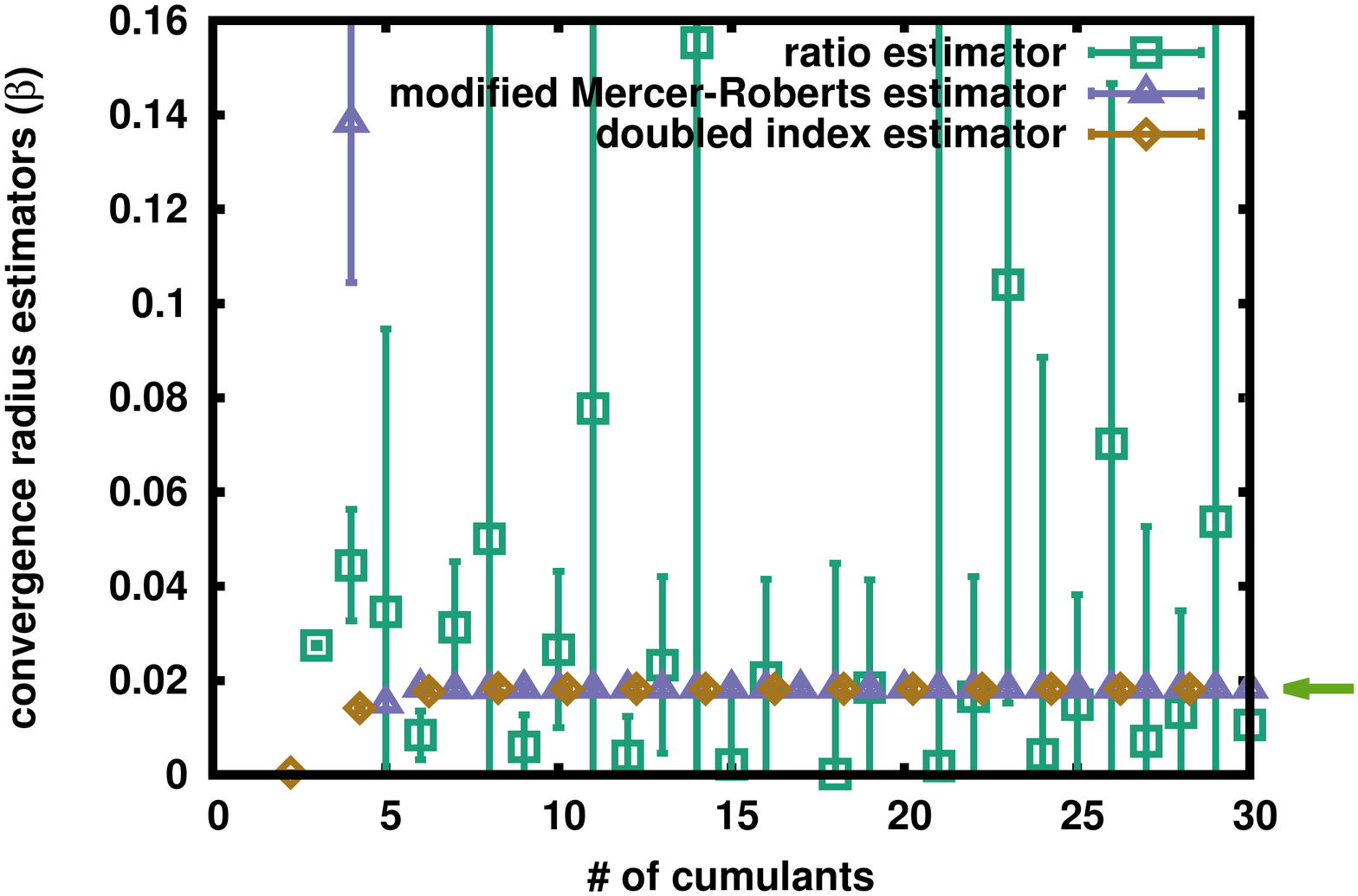}
  \includegraphics[width=0.5\textwidth]{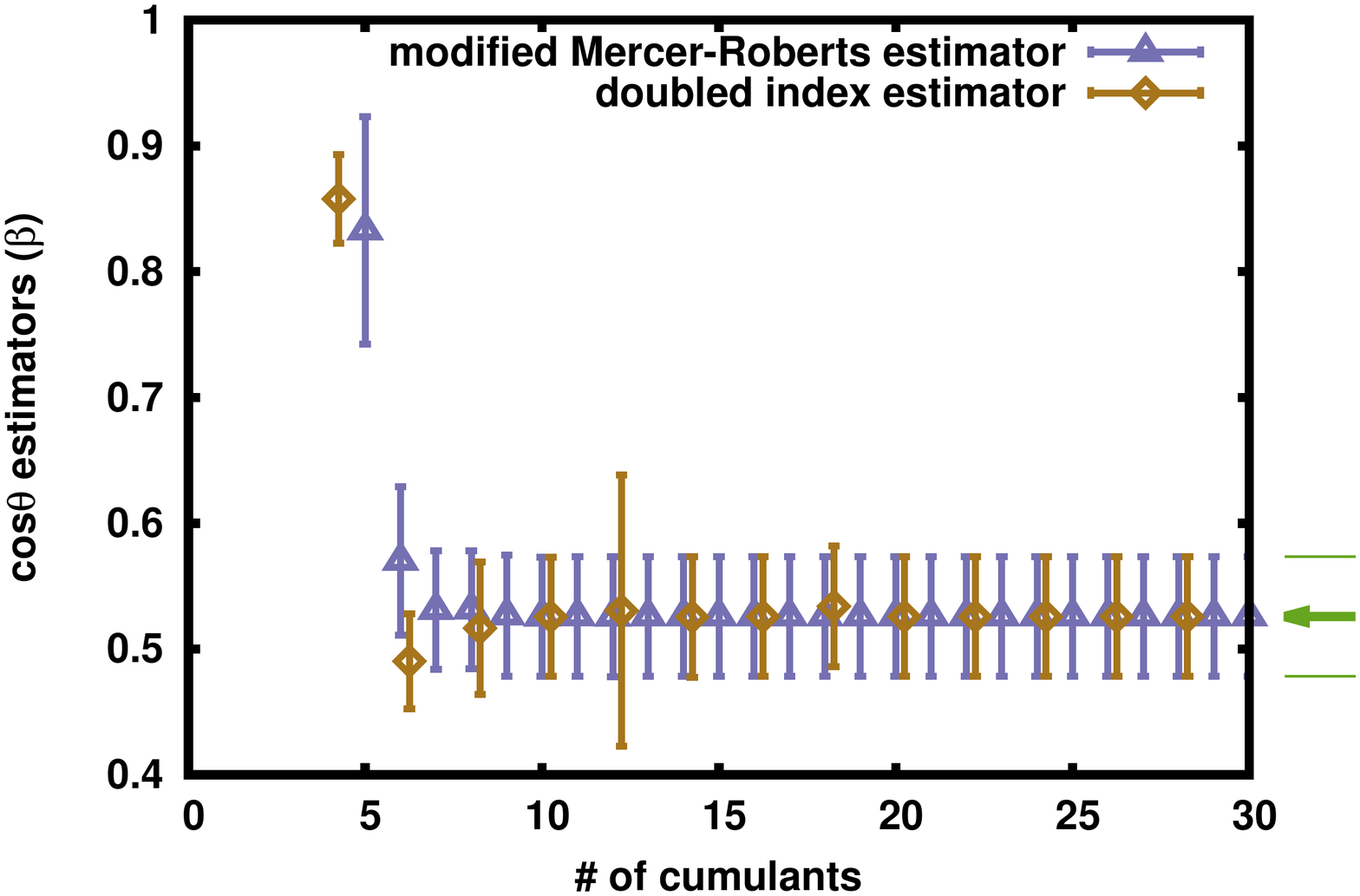}
  \caption{Top: The ratio, modified Mercer-Roberts and doubled index
    estimators for the distance of the leading pair of Fisher zeros, 
    compared to the result obtained from reweighting (the light green
    arrow). Data points are slightly shifted for visual clarity.
    Bottom: The modified Mercer-Roberts and doubled index estimators
    for the phase.
		}
  \label{fig:Fisher}
\end{figure}

Finally, let us turn to the analysis of the Fisher zeros. In this case we
do not calculate the cumulants starting from the partition function zeros,
and therefore the large cancellations of the errors, coming from the
correlations between the different cumulants, can be directly
demonstrated. This is shown in Fig.~\ref{fig:Fisher}, where we show
some estimators for the radius and for the cosine of the phase,  
and compare them to results obtained by reweighting 
to complex $\beta$. 
We can see that the ratio estimator again does not
work, while our newly proposed estimators work well, and give error bars 
identical to those obtained with reweighting.

\section{Summary and outlook}
\label{sec:concl}

In studies of QCD at finite chemical potential, it is a standard
approach to try to infer the position of the critical endpoint from
the radius of convergence of the Taylor expansion of the pressure,
either in the chemical potential over temperature, $\hat{\mu}$, or in
the fugacity parameter $\zeta=e^{\hat{\mu}}-1$. In this paper we have
demonstrated, on general grounds, that the simple ratio estimator 
cannot work if one wants to determine the radius of convergence in a
finite volume first, and take the thermodynamic limit afterwards, and
has serious drawbacks even if the radius of convergence is computed directly
in the thermodynamic limit. This follows from the analytic structure
of the grand canonical QCD partition function in a finite volume. 

The analytic structure of the partition function is also the starting
point for the construction of three new  
convergence radius estimators, that are designed precisely to work
well in a finite volume, where they determine the distance from the
origin of the closest Lee-Yang zero. We also constructed estimators
for the cosine of the phase of the leading Lee-Yang zero.
In this way we are able, at least in principle, to
locate the leading Lee-Yang zero in the complex $\hat{\mu}$ or $\zeta$
plane.  On a conceptual level, this provides a link between the
reweighting and Taylor expansion methods, demonstrating how to obtain
the leading Lee-Yang zero from the Taylor expansion, which was a
quantity that was traditionally obtained by reweighting in the
literature. 

Since the expansion order that can be reached is in practice limited,
and it is not known a priori how fast the estimators converge to
their asymptotic value, having different estimators at one's disposal
allows to cross-check whether for the highest accessible expansion
order the Taylor series is already dominated by a single
Lee-Yang zero or not. Interestingly, the relative statistical error of
our estimators at high enough order is expected to converge to a
finite value, contrary to what happens to the error of the expansion
coefficients themselves. This allows in principle to give a reliable
estimate of the correct location of the leading Lee-Yang zero.
In particular, our derivation makes it clear that 
for any given ensemble, if one uses our improved estimators, the study of the
Taylor coefficients and the direct determination of the Lee-Yang zeros
via reweighting should give identical results.

In order to test our proposals, we have performed an exploratory study
with $N_f=4$ unimproved staggered fermions on a small $6^3 \times 4$
volume, with bare quark mass chosen in order to have a first-order
phase transition (in the large-volume limit at fixed temporal extent in
lattice units), and for a temperature in the chirally-broken phase
close to the transition. For such a small system it is possible to
determine all the Lee-Yang zeros by standard methods, and thus to
obtain explicitly the radius of convergence and all the Taylor
coefficients. Our estimators work well, as expected,
achieving a rather accurate determination of the radius of convergence
with $O(15)$ Taylor coefficients in the case of $\zeta$, and only
around 6 coefficients in the case of $\hat{\mu}$.
This is not far from the situation currently available for realistic
lattices, i.e., lattices near the continuum limit of $N_f=2+1$ flavor
QCD with physical masses, where currently we have coefficients at most
up to $c_4$, corresponding to $\chi_8$ in the usual notation for 
the Taylor series in QCD, or $b_4$ with the notation
of~\cite{Vovchenko:2017xad}, which refers to the Fourier series in
imaginary chemical potential. While the number of Taylor
coefficients needed for an accurate determination of the radius of
convergence with our estimators is most certainly model dependent, and
therefore could be larger for realistic lattices, nevertheless one 
can hope that it still remains reasonable. At the same time, there is 
no reason to expect the situation to get better for the ratio estimator, 
meaning that the existing results of the first few orders of the ratio 
estimator~\cite{DElia:2016jqh,Bazavov:2017dus,Fodor:2018wul} are 
likely not that informative about the true radius of convergence.

A particularly encouraging aspect of our study is the statistical
accuracy of our estimators, which is vastly better than that of the
coefficients themselves. This can be explained by noticing that the
statistical errors on the coefficients are strongly correlated, which
leads to cancellations when they are combined into our convergence
radius estimators. 

We have also demonstrated our method on Fisher zeros. While this is
phenomenologically less interesting, it demonstrates explicitly the 
big cancellations of errors that we expect from our picture. In fact,
the results obtained with our new estimators are consistent with, and
have the same relative precision as those obtained independently 
via reweighting from the same ensemble.

There is a somewhat uplifting message to all this. While in our 
exploratory study we find over $100\%$ error bars on most of the
coefficients, it nevertheless turns out that with an ensemble of
reasonable size it is possible to determine the position of the
leading Lee-Yang zero quite accurately. In other words, the errors on
the Taylor coefficient do not reflect the actual uncertainty with
which the leading Lee-Yang zero can be obtained.
It might then even be possible to extract the position of the leading
Lee-Yang zero already from existing ensembles, just not with the
method currently pursued in the literature (i.e., ratio estimators
from Taylor expansion coefficients). 
Determining an efficient method for this is an important task for the
future. 

\section*{Acknowledgements}
We thank S. Katz, D. Nográdi and  Sz. Borsányi for useful discussions.
This work was partially supported by the Hungarian National
Research, Development and Innovation Office - NKFIH grant KKP126769 
and by OTKA under the grant OTKA-K-113034.

\end{document}